\documentclass[11pt]{article}
\usepackage[]{acl}
\usepackage{times}
\usepackage{latexsym}
\usepackage[T1]{fontenc}
\usepackage[utf8]{inputenc}
\usepackage{microtype}
\usepackage{graphicx}
\usepackage{booktabs}
\usepackage{makecell}
\usepackage{amssymb}
\usepackage{xcolor}
\usepackage{pifont}
\usepackage{longtable}
\usepackage{array}
\usepackage{tcolorbox}
\tcbuselibrary{listings,breakable,skins}
\usepackage{stfloats}
\usepackage{xurl}
\newcommand{\cmark}{\textcolor{green!60!black}{\ding{51}}}
\newcommand{\xmark}{\textcolor{red}{\ding{55}}}
\newcolumntype{L}[1]{>{\raggedright\arraybackslash}p{#1}}

\title{SEABED: SouthEast Asian Benchmark for Evaluating Audio Reasoning}

\author{
Harshit Rajgarhia, Asif Shaik, Rachuri Lokesh, Sushanta Kumar Pani, Abhishek Mukherji \\
Centific AI Research \\
\texttt{harshit.rajgarhia@centific.com}
}

\begin{document}
\maketitle

%==================================================================
\begin{abstract}
Modern audio-language models are no longer judged only on what words they can
transcribe, but on whether they can \emph{reason} over what they hear: recovering
meaning that lives in tone and prosody, telling dialects and regional languages
apart, and resolving ambiguity that the written form leaves open. This capability
is now measured by a growing family of audio-reasoning benchmarks, but almost
entirely in English and on general-domain audio. Southeast Asia (SEA)
% , 
% % home to
% % over $650$ million speakers of more than a thousand tonal and prosody-rich
% % languages,
is served instead by benchmarks that inherit an English task taxonomy
of recognition, translation, and paralinguistic classification, and therefore
test whether a model \emph{hears} SEA speech rather than whether it can reason
from it. We introduce \textbf{SEABED}, an \emph{audio-first} question answering
dataset designed to benchmark language and audio reasoning models on SEA speech. SEABED comprises
a suite of six audio-reasoning tasks built \emph{entirely from real, openly
available SEA speech corpora}, yielding $5{,}404$ question-answer pairs. We evaluate six frontier and region-specific
audio LLMs: even the state-of-the-art model Gemini~3.5~Flash achieves only
\textbf{50.3\%} weighted average accuracy. \iffalse \textcolor{red}{SEABED measures not just whether models answer, but whether they listen and reason.} \fi SEABED evaluates not only answer accuracy, but also whether models' stated reasoning is grounded in the audio evidence. A sample of the benchmark data is available here:
\url{https://huggingface.co/datasets/CentificAIResearch/SEABED}.
\end{abstract}

%==================================================================
\section{Introduction}

\begin{figure*}[t]
\centering
\includegraphics[width=\textwidth]{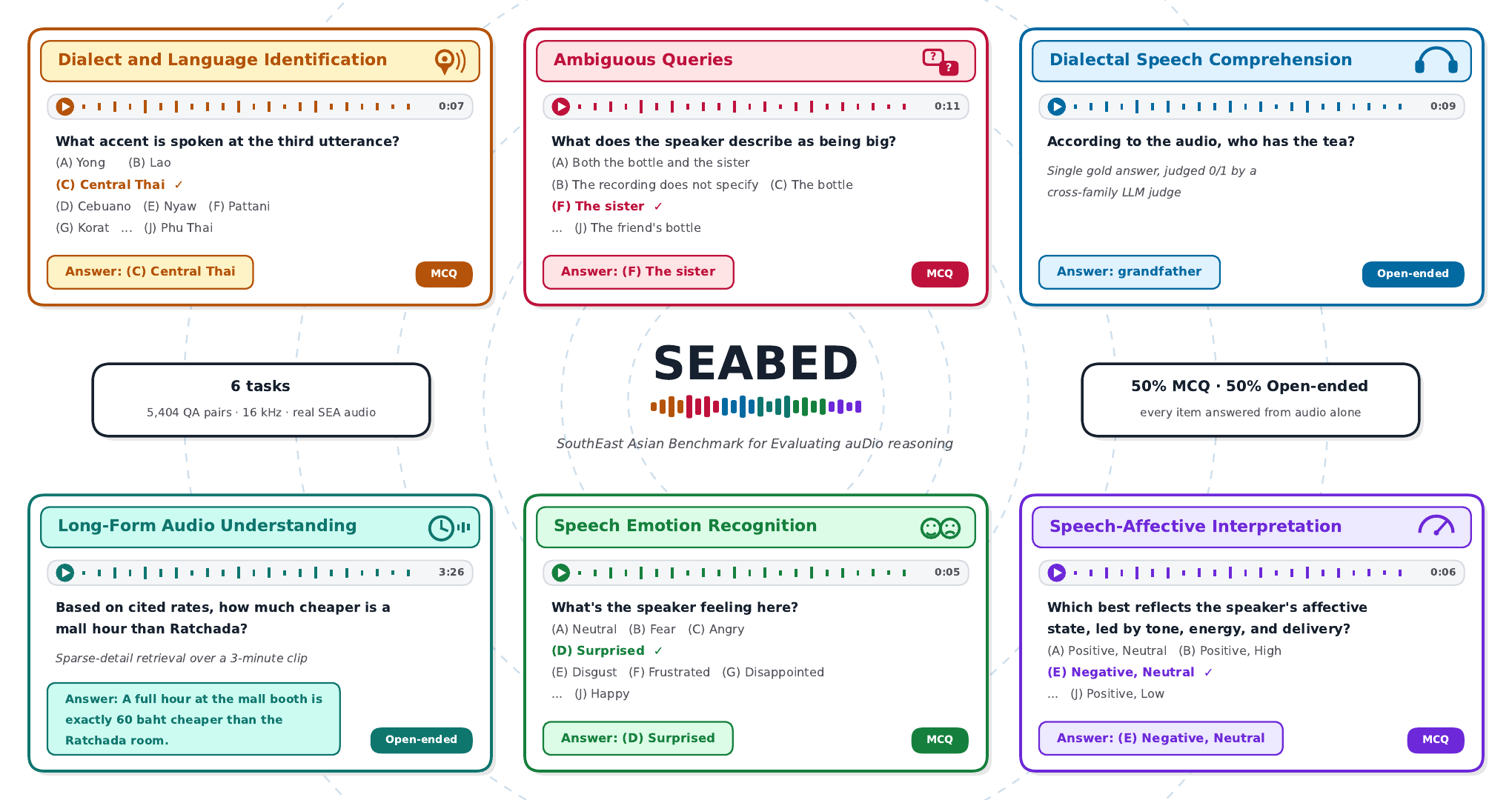}
\caption{SEABED at a glance: one example item per task (question, options, gold answer).}
\label{fig:teaser}
\end{figure*}

%*********************************
%paragraph 1 to 4 compressed 
%*********************************
% Audio-language models are increasingly judged not on transcription but on
% \emph{audio reasoning}: drawing, from cues such as pitch, timing, voice quality,
% and prosody, conclusions the words alone cannot support. Telling agreement from
% refusal under identical wording, reading affect from delivery, or resolving a
% syntactic ambiguity from its prosody are inferences of this kind, and none
% survives reduction to a transcript.

Spoken-language systems have long been judged by transcription accuracy, but
audio-language models are built for \emph{audio reasoning}, drawing on cues such
as pitch, timing, voice quality, and prosody to reach conclusions the words alone
cannot support. Distinguishing agreement from refusal under identical wording,
reading affect from delivery, or resolving a syntactic ambiguity from its prosody
all depend on how something is said, not just what is said, and none survives
reduction to a transcript.

Nowhere is this more consequential than in Southeast Asia, with over $650$
million people and more than a thousand indigenous languages \citep{seacrowd}.
Many are tonal, so a pitch contour alone can separate words
\citep{tonefunctionalload}, and dialect continua share spelling while identity,
register, and often meaning live in prosody, so a transcript can miss the
message. Region-specific models are already deployed, SeaLLMs-Audio
\citep{seallmsaudio} and MERaLiON-AudioLLM \citep{meralion}, yet no SEA suite
tests whether they reason over these acoustic phenomena rather than merely hear
them.
 
Closing this gap takes more than translating an English benchmark. General-domain
reasoning suites \citep{airbench,audiobench,mmau,mmar,mmsu,mmaupro} are English
and general-domain and carry a validity flaw: models answer nearly half of MMAU
from text priors with the audio silenced \citep{audiomcq}, so accuracy alone
cannot certify listening. SEA suites \citep{seallmsaudio,seaspeechbench} add
languages but keep a recognition-and-classification taxonomy that shows
underperformance without testing reasoning, as also seen for Indic
\citep{indicvoices} and Persian \citep{parsabench} speech. A SEA reasoning
benchmark must therefore control the language, the phenomena, and the evidence
that the model actually listened.
 
We introduce \textbf{SEABED}, the SouthEast Asian Benchmark for Evaluating Audio
Reasoning: an \emph{audio-first} dataset of $5{,}404$ pairs whose six tasks each
isolate one phenomenon SEA speech makes acoustic (Figure~\ref{fig:teaser}), built
from real, openly available corpora. \iffalse \textcolor{red}{Across six frontier and region-specific
audio LLMs the benchmark is unsolved, the best resolves barely half, and a
reasoning-grounding audit shows accuracy and genuine listening diverge: for the
regional model, one correct answer in four is ungrounded. SEABED thus separates
models that listen and reason from those that merely answer.}\fi Across six frontier and region-specific audio LLMs the benchmark is unsolved,
the best resolves barely half, and a reasoning-grounding audit shows that
answer accuracy and grounding quality can diverge: for the regional model,
one correct answer in four is accompanied by ungrounded stated reasoning.
SEABED thus separates answer accuracy from the grounding quality of models'
reported reasoning. Our contributions are four-fold.

\begin{itemize}\itemsep0.2em
  \item \textbf{A reasoning-first, audio-first SEA benchmark.} Six tasks over
  SEA-specific acoustic phenomena; $5{,}404$ QA pairs answered from audio alone,
  evenly split between multiple-choice and open-ended.
\item \textbf{A listen-and-reason audit protocol.}
A model-agnostic, two-axis diagnostic that reads each model's stated reasoning
against the audio, assesses its grounding in the audio evidence, and types
every error.
  \item \textbf{Construction from real, open SEA audio.} Built entirely from
  existing open corpora with no new recording and \emph{no synthesized speech}.
  \item 
  \textbf{A quantified gap between accuracy and reasoning grounding.}
Across six models the best reaches only 50.3\%; up to a quarter of the audited
models' correct answers have ungrounded stated reasoning, and failures are
comprehension-dominated.

\end{itemize}

%******************

%==================================================================
\section{Related Work}

\paragraph{From audio recognition to audio reasoning.}
AIR-Bench \citep{airbench} and AudioBench \citep{audiobench} established
broad audio understanding and instruction following; a reasoning wave
followed, with MMAU spanning $27$ skills over $10$k clips \citep{mmau}, MMAR
adding hierarchically layered questions with chain-of-thought rationales
\citep{mmar}, MMSU grounding $47$ tasks in linguistic theory \citep{mmsu},
and MMAU-Pro broadening to $49$ skills with in-the-wild audio and a mixed
multiple-choice and open-ended format \citep{mmaupro}. These benchmarks
refined the design values SEABED adopts, namely deliberate multi-hop items,
careful distractors, and mixed formats, but all treat English general-domain
audio as the default, and language is never a controlled variable. Like
MedMosaic \citep{medmosaic}, we carry these values into a domain where data
is scarce and reasoning is the point, here the languages of Southeast Asia.

\paragraph{Does the model actually listen?}
A parallel line questions whether LALMs use the audio at all: \citet{audiomcq}
document a \emph{zero audio-contribution} phenomenon, with models answering
correctly from text alone on $49.8\%$ of MMAU items even when the audio is
silenced, and respond with audio-contribution-aware post-training. \iffalse \textcolor{red}{SEABED
attacks the same threat from the benchmark side: distractors are anchored to
audible confusability rather than topic, and the reasoning-grounding audit
(\S\ref{sec:analysis}) classifies each model's stated reasoning as grounded,
partial, ungrounded, or contradictory, turning \emph{did the model listen}
from an assumption into a measured quantity.}\fi SEABED attacks the same threat from the benchmark side:
distractors are anchored to audible confusability rather than topic, and the
reasoning-grounding audit (\S\ref{sec:analysis}) classifies each model's stated
reasoning as grounded, partial, ungrounded, or contradictory, providing a
diagnostic of whether the model's stated reasoning is supported by the audio
evidence.

\paragraph{Frontier and region-specific audio LLMs.}
End-to-end LALMs pair an audio encoder with a language model and answer
directly from sound, and dedicated reasoning variants such as Audio-Reasoner
train on large-scale audio chain-of-thought data \citep{audioreasoner}. For
Southeast Asia, region-specific models have emerged: SeaLLMs-Audio covers
Indonesian, Thai, and Vietnamese \citep{seallmsaudio}, and MERaLiON-AudioLLM
targets Singapore's multilingual setting, including Singlish \citep{meralion}.
SEABED evaluates both families side by side, asking whether regional
specialization buys regional listening.

\paragraph{SEA speech resources and evaluation.}
SEACrowd standardizes corpora across nearly a thousand SEA languages, and its
speech holdings are overwhelmingly ASR and TTS \citep{seacrowd}: the raw
material of recognition, not reasoning. The evaluation suites built on this
landscape follow suit. SeaBench-Audio accompanies SeaLLMs-Audio
\citep{seallmsaudio}, and SEA-SpeechBench spans eleven SEA languages with
over $97$k samples \citep{seaspeechbench}, yet both organize their items into
recognition, translation, and paralinguistic classification, the taxonomy our
introduction argues tests hearing rather than reasoning. Closest in spirit to
our affective tasks, CPQA introduces contextual paralinguistic question
answering with human verification \citep{cpqa}; we adapt its design to SEA
languages under a strict audio-first protocol. What none of these provide,
and what SEABED contributes, is a benchmark whose every item demands an
inference that SEA speech makes acoustic: dialect and language identity
carried purely by pronunciation, sentence readings that only prosody settles,
and dialectal speech a standard transcript would smooth away.

%==================================================================
\section{The SEABED Benchmark}
\label{sec:benchmark}

\begin{figure*}[t]
\centering
\includegraphics[width=\textwidth]{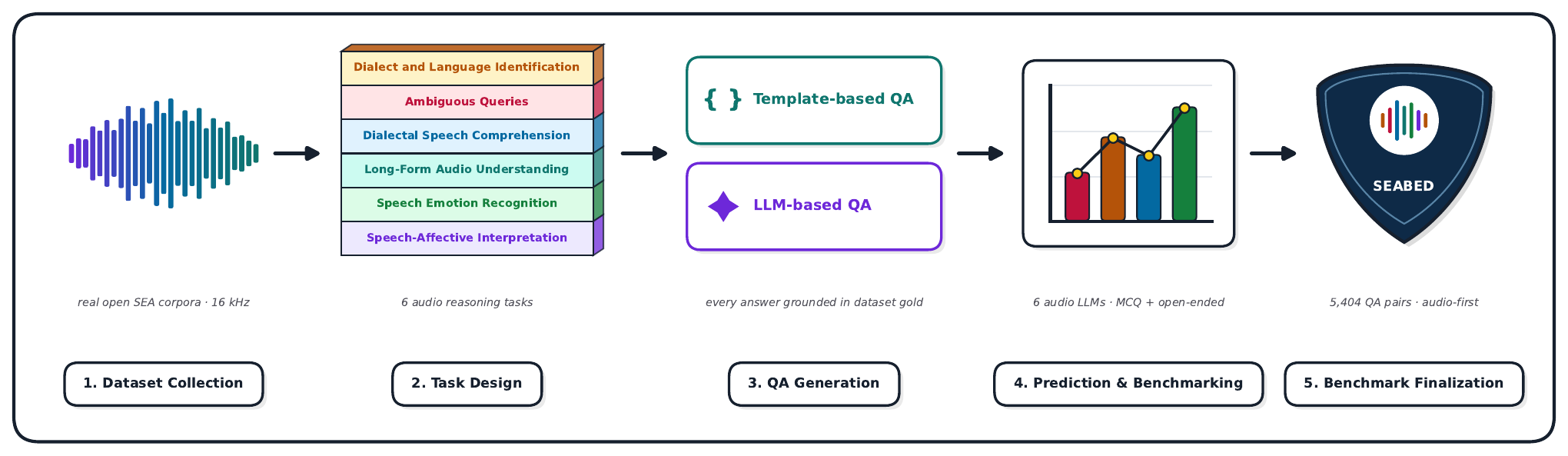}
\caption{SEABED construction and evaluation pipeline.}
\label{fig:pipeline}
\end{figure*}

SEABED is defined less by a list of tasks than by a set of principles that
every admitted item satisfies; this section states those principles, how the
source data was found, what the six tasks measure, and how every item was
generated and checked.

%trimmed and appended to appendix
\subsection{Design Principles}
Four principles govern every task.
\begin{itemize}\itemsep0.2em
    \item \textbf{Audio-first.} At test time the model receives only the audio
    clip and the question; transcripts, metadata, and labels are used in
    construction only, never exposed.
    \item \textbf{Reasoning over recognition.} Every task requires an inference
    the transcript alone cannot settle, which we verify with a transcript
    ablation (\S\ref{sec:ablations}) and reasoning-grounding audit
    (\S\ref{sec:analysis}).
    \item \textbf{Real audio only.} All clips are curated from existing, openly
    licensed SEA corpora and resampled to $16$\,kHz; nothing is newly recorded
    and no synthesized speech is admitted.
    \item \textbf{Guess-resistant by construction.} Multiple-choice items use up
    to ten confusability-anchored options (following MMAU-Pro and MedMosaic
    \citep{mmaupro,medmosaic}); open-ended items are scored against a single gold
    answer, and gold positions are debiased so no answer slot leaks information
    \citep{audiomcq}.
\end{itemize}

\subsection{Datasets}
\label{sec:datasetdesign}

SEABED draws on ten openly licensed corpora spanning Indonesian, Thai, and
Malay (Appendix~\ref{app:datasets}). Six corpora underpin the structural
and comprehension tasks: INDspeech, parallel news speech read in four
Indonesian accents; the SLSCU Thai Dialect Corpus, regional-dialect speech
with dialect transcripts; STRUCT\_AMB\_IND, structurally ambiguous
Indonesian sentences each recorded under an instructed reading; and
LOTUSDIS, ASR-IndoCSC, and ASR-MalCSC, spontaneous multi-party
conversations of seven to twenty-three minutes. The remaining four provide
emotion-annotated speech for the affect tasks: THAI SER, E-SERAVD,
IndoWaveSentiment, and SeaBench-Audio.

\subsection{Task Suite}
\label{sec:tasks}

SEABED comprises six tasks (Table~\ref{tab:tasks},
Appendix~\ref{app:stats}): three built on SEA-specific structural phenomena (\emph{dialect and language
identification}, \emph{prosodic ambiguity resolution}, \emph{dialectal
speech comprehension}), one stressing extended context (\emph{long-form
audio reasoning}), and two paralinguistic affect tasks (\emph{speech emotion
recognition}, \emph{speech-affective interpretation}). Five tasks contribute
$1{,}008$ items each; long-form contributes $364$, capped by the scarcity of
long, openly licensed SEA recordings, for $5{,}404$ question-answer pairs in
total, split evenly between multiple-choice and open-ended within every
task; the full per-task composition, with languages and audio hours, is
given in Table~\ref{tab:taskstats} (Appendix~\ref{app:stats}).

% --- keep your existing tab:tasks table here unchanged ---

\subsubsection{Dialect and Language Identification}
Each stimulus concatenates four utterances, one per lect, in random order
with silence gaps: Batak, Javanese, Sundanese, and Standard for Indonesian;
Central, Khummuang, Korat, and Pattani for Thai. Content is controlled or
uninformative (the Indonesian half is parallel; the Thai lects share no
sentences), so identification must come from phonology alone. Reasoning,
not classification, is the point: the model must segment the stream,
identify each lect, hold the sequence, and answer four question variants
over it (naming one segment's lect, recovering the full order, locating a
named lect, comparative queries), so a correct answer requires relational
inference over the audio timeline rather than a single label.

\subsubsection{Prosodic Ambiguity Resolution}
Every item is a structurally ambiguous Indonesian sentence with exactly two
grammatically valid readings; the speaker was instructed to realize one,
fixing the intended reading at recording time. A text-only system is
provably at chance: answering requires reasoning from prosodic evidence
(pause placement, phrase grouping, stress) to syntactic structure, an
inference no transcript supports. Three variants form a ladder of reasoning
depth: single-utterance disambiguation; dual renditions by the same
speaker, one per reading, whose meanings must both be identified in order;
and a targeted variant that must isolate one rendition while a
word-identical competitor plays in the same clip.

\subsubsection{Dialectal Speech Comprehension}
The model answers content questions over Thai regional-dialect speech
spanning Pattani Malay, Korat, and Khummuang, where non-standard phonology,
reduced forms, and region-specific lexis obscure content that citation
speech would make trivial. The reasoning demand is comprehension under
dialect shift: the model must resolve unfamiliar surface forms to their
intended referents (numerals, person and kin references) from partial
acoustic evidence, not merely transcribe them. Curation concentrates on the
material where this inference is hardest, golds are anchored to the dialect
transcript, the only faithful record of the audio, and audio dependence is
probed explicitly in the transcript ablation (\S\ref{sec:ablations}).

\subsubsection{Long-Form Audio Reasoning}
The model listens to an entire spontaneous multi-speaker conversation,
seven to twenty-three minutes long: Thai office meetings and Indonesian and
Malay telephone conversations. Questions are built so the answer is never
stated in any single segment: answer must be \emph{derived} by locating and
combining evidence from distant parts of the recording, through long-range
recall, cross-segment integration, multi-hop inference, content-based
speaker attribution and implicit inference. Reasoning is over spoken content only; no prosodic or
timestamp-based questions are asked.

\subsubsection{Speech Emotion Recognition}
The model assigns one discrete emotion from a fixed ten-way set (angry,
disappointed, disgust, fear, frustrated, happy, neutral, sad, surprised, excited)
over Thai and Indonesian emotional speech; ground truth is the source
corpus's canonical label. The reasoning lies in judging delivery against
words: the lexical content rarely names the feeling, so the model must
infer it from tone, pacing, and energy rather than sentiment keywords.

\subsubsection{Speech-Affective Interpretation}
SAI asks for the speaker's position on Russell's circumplex: a joint
judgment of valence (positive, negative, neutral) and arousal (high, low,
neutral). This is structured affective reasoning rather than labeling:
prosody must be decomposed into two independent dimensions and both must be
correct for an item to count. The four attested cells are exactly balanced
at $252$ items each.

\subsection{QA Generation}
\label{sec:construction}

All items flow through a shared pipeline (Figure~\ref{fig:pipeline}) with
two generation routes, chosen by what each source corpus can guarantee. For
four tasks (DIALECT, AMBIGUOUS, SER, SAI) the gold is derived
deterministically from the source corpus: accent and lect labels, the
reading a speaker was instructed to realize, canonical emotion labels, and
valence-arousal cells; no model ever authors an answer, and questions are
instantiated from templates (DIALECT, SER, SAI) or LLM-drafted against the
fixed gold (AMBIGUOUS). On these fixed golds we engineer the option sets:
in-language near-neighbor lects and same-family emotions rather than
eliminable foils, permutation and exact-reversal traps on order questions,
and safe-play abstention options (``cannot be determined from the audio'')
that bait hedging models and are never correct, with options shuffled and
gold positions debiased. For the remaining two tasks (DSC, LONG\_FORM), no
source annotation yields question-answer pairs directly, so both are
authored by Gemini~3.1~Pro \citep{gemini31pro}, conditioned on the audio
with its aligned gold transcript and metadata, anchoring every question to
what is actually said; these answers are scored by a cross-family judge
(\S\ref{sec:setup}), so no model grades its own generations.
Generated questions may not quote, translate, or closely paraphrase the
transcript; every build passes automated checks on item counts, language
and gender balance, option integrity, and text-only answerability; and all
sampling is seeded, so the benchmark regenerates deterministically. \iffalse \textcolor{red}{With four tasks on source-annotation gold and the generator-authored two transcript-anchored, leakage-checked, and cross-family judged, SEABED needs no per-item expert pass; human validation is left to future work (\S\ref{sec:conclusion}).} \fi With four tasks based on source-annotation gold and the two generator-authored tasks transcript-anchored, leakage-checked, and cross-family judged, we apply source-grounded and automated checks throughout construction; systematic human validation remains an important direction for future work
(\S\ref{sec:conclusion}).

%==================================================================
\section{Experiments}

\begin{table*}[t]
  \centering
  \caption{Accuracy on SEABED (\%). Task codes follow Table~\ref{tab:tasks}:
  DIALECT (dialect and language identification), AMBIG (prosodic ambiguity
  resolution), DSC (dialectal speech comprehension), LONG (long-form audio
  reasoning), ER (speech emotion recognition), SAI (speech-affective
  interpretation). Best per column in \textbf{bold}. Weighted Avg is the
  sample-weighted mean over all items.}
  \label{tab:sea_audio_benchmark}
  \resizebox{\textwidth}{!}{%
  \begin{tabular}{l rrrrrr rrrrrr r}
    \toprule
    & \multicolumn{6}{c}{Multiple-choice} & \multicolumn{6}{c}{Open-ended} & \\
    \cmidrule(lr){2-7}\cmidrule(lr){8-13}
    Model & DIALECT & AMBIG & DSC & LONG & ER & SAI
          & DIALECT & AMBIG & DSC & LONG & ER & SAI & \makecell{Weighted\\Avg} \\
    \midrule
    Gemini 2.5 Pro   & 37.5 & 47.6 & 68.1 & 76.1 & 30.8 & 36.7 & 29.1 & 35.9 & 44.3 & 64.6 & 31.7 & 30.8 & 41.6 \\
    Gemini 3.5 Flash & \textbf{41.5} & 60.5 & \textbf{69.4} & \textbf{80.2} & \textbf{49.4} & \textbf{45.8} & \textbf{34.1} & 48.5 & \textbf{48.7} & 66.7 & \textbf{48.8} & \textbf{37.7} & \textbf{50.3} \\
    Gemini 3.6 Flash & 39.3 & \textbf{60.7} & 68.3 & 79.1 & 40.3 & 42.7 & 32.5 & \textbf{49.4} & 42.5 & \textbf{68.3} & 41.2 & 36.6 & 47.4 \\
    GPT-Audio 1.5    & 18.7 & 38.9 & 47.0 & 50.5 & 29.4 & 31.3 & 22.7 & 31.3 & 21.9 & 50.2 & 34.9 & 29.8 & 32.0 \\
    MERaLiON-3-10B   & 17.9 & 38.3 & 42.7 & 20.3 & 30.0 & 38.7 & 19.5 & 27.8 & 28.4 & 12.9 & 32.8 & 29.0 & 29.4 \\
    SeaLLMs-Audio-7B & 10.6 & 26.8 & 11.6 & 15.9 & 18.5 & 16.3 & 14.0 & 27.0 & 12.0 & 11.3 & 16.9 & 25.0 & 17.5 \\
    \bottomrule
  \end{tabular}}
\end{table*}

\subsection{Experimental Setup}
\label{sec:setup}

\paragraph{Models.}
We benchmark six audio LLMs spanning frontier and region-specific
families: four hosted frontier models (Gemini~2.5~Pro, Gemini~3.5~Flash,
Gemini~3.6~Flash, GPT-Audio~1.5) and two open region-specific models run
locally (MERaLiON-3-10B, SeaLLMs-Audio-7B). Question generation uses
Gemini~3.1~Pro Preview; open-ended answers are judged by the cross-family
Grok~4.5; and the reasoning-trace audit (\S\ref{sec:analysis}) is
classified by Claude~Opus~4.8, also from a family not under evaluation, so
no benchmarked family grades its own reasoning. \iffalse Question generation uses
Gemini~3.1~Pro Preview; open-ended answers are judged by the cross-family
Grok~4.5; and the reasoning-trace audit (\S\ref{sec:analysis}) is
classified by Claude~Opus~4.8, also from a family not under evaluation, so
no benchmarked family grades its own reasoning.\fi This cross-family design
avoids direct self-grading, but it does not substitute for systematic human
validation. The full roster, with
roles and citations, is given in Table~\ref{tab:roster}
(Appendix~\ref{app:models}).

\paragraph{Evaluation.}
Multiple-choice items are scored by exact match against the shuffled gold
option. Open-ended items are scored by the Grok~4.5 judge: binary
correct/incorrect for five tasks (SAI requires both valence and arousal),
and a graded $0$ to $1$ rubric weighting correctness, relevance, and
completeness for long-form's longer answers. We report per-task accuracy
per format and, as the headline number, the \emph{Weighted Avg}, the
sample-weighted mean over all items. All audio is presented at $16$\,kHz;
all generation, judging, and audit prompts appear verbatim in
Appendix~\ref{app:prompts}.

\subsection{Results}
\label{sec:results}

Table~\ref{tab:sea_audio_benchmark} reports accuracy for all six models over
the twelve task and format cells and the weighted average. Five patterns
stand out. \textbf{(1) SEABED is hard for every model.} The best system,
Gemini~3.5~Flash, reaches only $50.3\%$ weighted accuracy, and the two open
region-specific models trail far behind at $29.4\%$ (MERaLiON-3-10B) and
$17.5\%$ (SeaLLMs-Audio-7B, near chance on several tasks), mirroring the
frontier-versus-open gap of MMAU-Pro and MedMosaic \citep{mmaupro,medmosaic}:
region-specific training has not yet produced reasoning over SEA speech.
\textbf{(2) The frontier Flash models lead almost everywhere.}
Gemini~3.5~Flash is best in nine of twelve cells and on the average, with
Gemini~3.6~Flash close behind; the larger Gemini~2.5~Pro trails both, so
recency and audio tuning matter more than raw scale. \textbf{(3) Tasks
separate sharply.} Long-form reasoning and DSC are most tractable for
frontier models (up to $80.2\%$ and $69.4\%$ multiple-choice), dialect and
language identification is hardest for every model (best $41.5\%$ MCQ,
$34.1\%$ open-ended), and the affect tasks sit between: naming a
fine-grained lect from a short clip is a demanding, high-cardinality
inference. \textbf{(4) Open-ended answering is markedly harder than
multiple choice.} Removing the option list lowers accuracy nearly
everywhere (DSC falls from $69.4\%$ to $48.7\%$ for the best model), an
inflation we quantify in \S\ref{sec:abl-options} and explain
mechanistically in \S\ref{sec:analysis}. \textbf{(5) Model provenance
explains the outliers.} MERaLiON-3-10B was evaluated by its authors on
audio up to roughly five minutes \citep{meralion}; SEABED's conversations
run seven to twenty-three, and its long-form cells collapse to $20.3$ and
$12.9$, its worst anywhere and the only case where transcripts beat audio
(\S\ref{sec:ablations}). Capability envelopes from training, not reasoning
alone, set the floor on long SEA audio.

\subsection{Reasoning-Grounding and Failure Analysis}
\label{sec:analysis}

Accuracy tells us \emph{whether} a model answered, not \emph{how}. We
therefore audit the reasoning behind every answer along two axes:
\emph{grounding} (\texttt{grounded}, \texttt{partial}, \texttt{ungrounded},
\texttt{contradictory}), labeled over all responses, and \emph{failure type}
(\texttt{perception}, \texttt{comprehension}, \texttt{hallucination},
\texttt{reasoning}, \texttt{abstention}, \texttt{format}), labeled over
wrong answers; full label definitions are in Table~\ref{tab:taxonomy}. \iffalse\textcolor{red}{A
correct answer with grounded reasoning is \emph{genuine audio reasoning};
one with ungrounded or contradictory reasoning is a \emph{lucky guess}.}\fi
For conciseness, we refer to a correct answer with grounded stated reasoning
as \emph{genuine audio reasoning} and one with ungrounded or contradictory
stated reasoning as a \emph{lucky guess}. These labels describe the grounding
of the model's reported rationale under our audit and do not establish that
the audio causally determined the prediction.

The audit covers a stratified $10\%$ sample of every task ($543$ items per
model; $1{,}629$ responses) for one model per family: Gemini~3.6~Flash,
GPT-Audio~1.5, and MERaLiON-3-10B. Claude~Opus~4.8 \citep{claudeopus48},
from a family not under evaluation, classifies each stated reasoning against
the reference transcript and gold metadata; an author spot check of a small
random subset agreed with its labels, though we claim no systematic human
review. {Figure~\ref{fig:radar}a maps the genuine rate by task,
Figure~\ref{fig:radar}b both axes per model, and Table~\ref{tab:genuine} the
headline split.

\begin{figure*}[t]
\centering
\begin{minipage}[c]{0.40\textwidth}
  \centering
  \includegraphics[width=\linewidth]{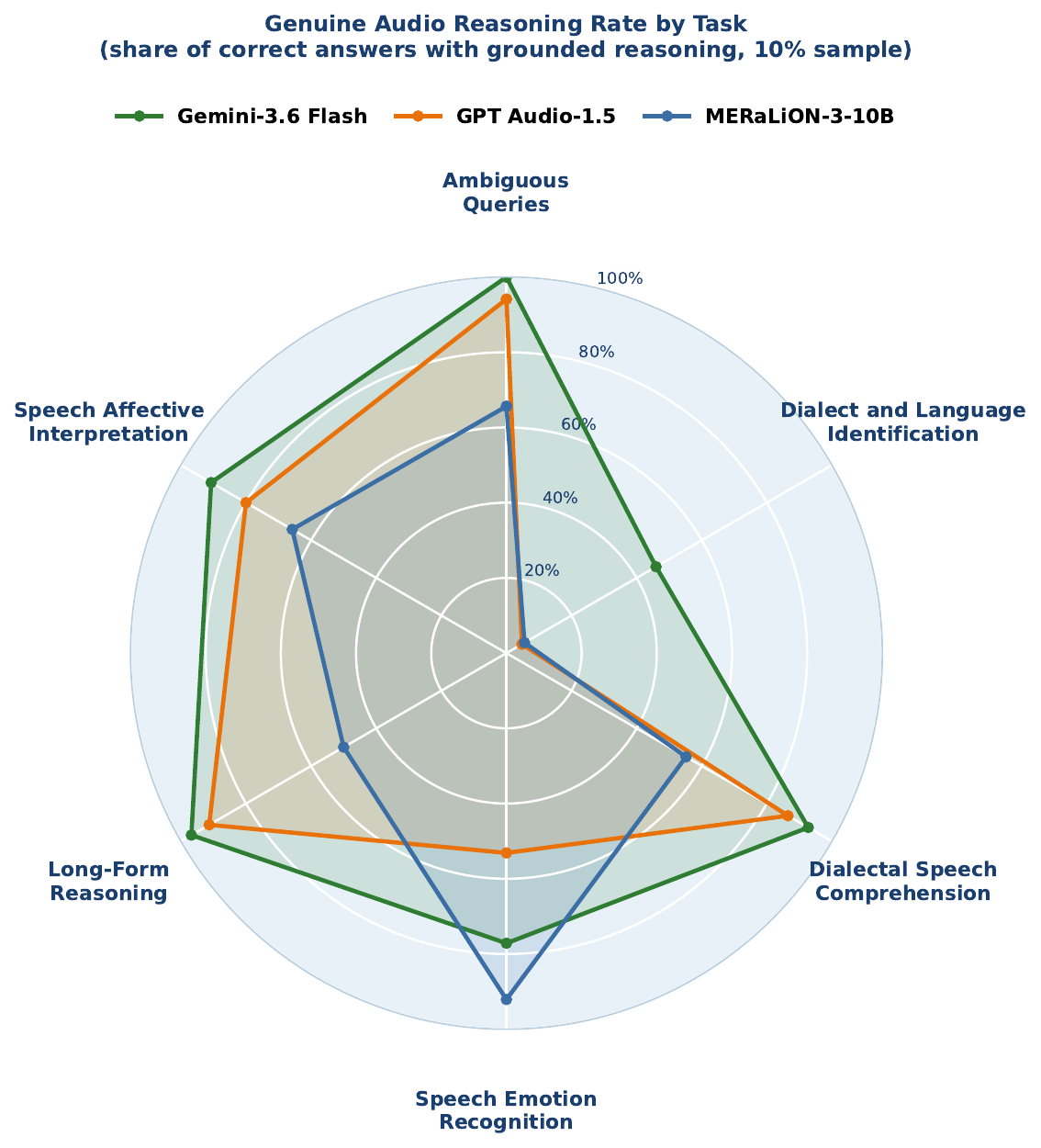}\\[-0.1em]
  {\small (a)}
\end{minipage}\hfill
\begin{minipage}[c]{0.58\textwidth}
  \centering
  \includegraphics[width=\linewidth]{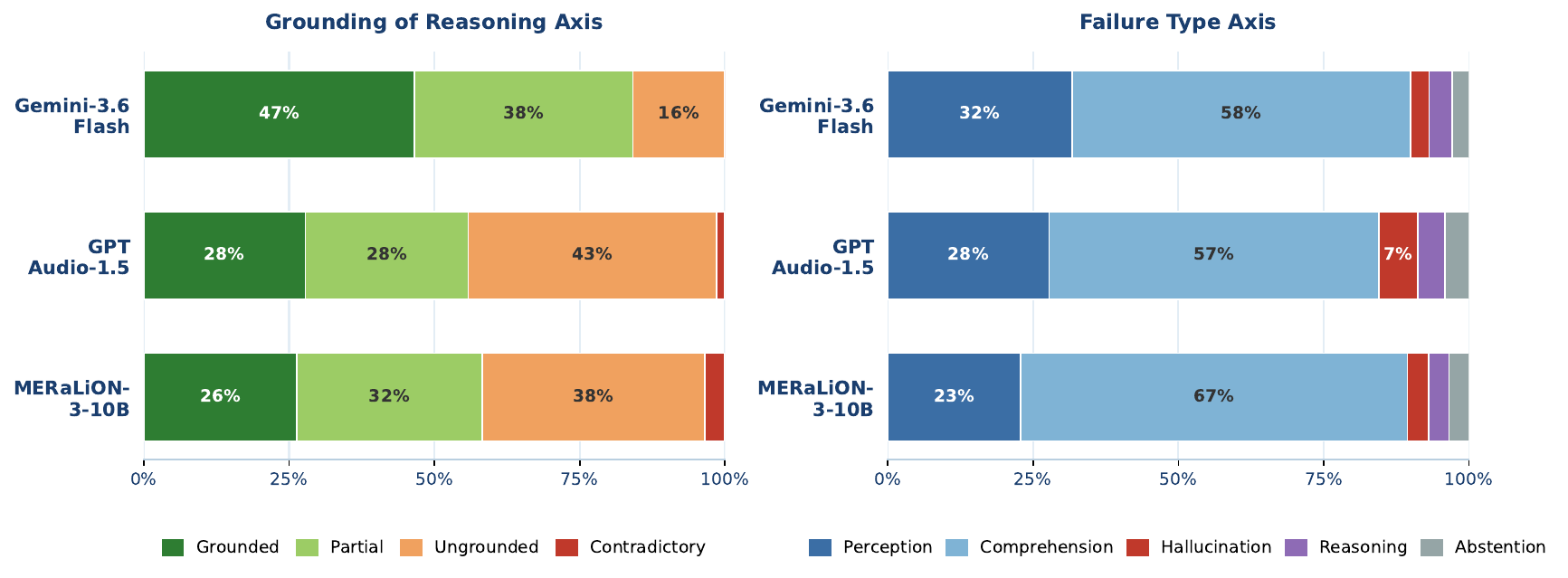}\\[0.3em]
  {\small (b)}
\end{minipage}
\caption{The reasoning-trace audit at a glance. \textbf{(a)} Genuine audio
reasoning rate by task: the share of each model's correct answers whose reasoning
is grounded in the audio. \textbf{(b)} The two audit axes per model: the grounding
of reasoning axis, a distribution over all responses, and the failure type axis, a
distribution over incorrect answers.}
\label{fig:radar}
\end{figure*}

\begin{table}[t]
\centering
\small
\resizebox{\columnwidth}{!}{%
\begin{tabular}{lrrrrr}
\toprule
Model & $n$ & Acc. & \makecell{Genuine \\(of corr.) } & \makecell{Partial\\(of corr.)} & \makecell{Lucky \\(of corr.)} \\
\midrule
Gemini 3.6 Flash & 543 & 48.8 & \textbf{86.0} & 5.3 & \textbf{8.7} \\
GPT-Audio 1.5    & 543 & 30.8 & 71.3 & 13.8 & 15.0 \\
MERaLiON-3-10B   & 543 & 28.9 & 63.1 & 10.8 & 26.1 \\
\bottomrule
\end{tabular}}
\caption{How correct answers were reached, on the audited $10\%$ sample
(percentages).}
\label{tab:genuine}
\end{table}

\iffalse \paragraph {1. Correct is not the same as earned.}\textcolor{red}{
Only $75.7\%$ of correct answers are earned by genuine audio reasoning;
$15.1\%$ are lucky guesses and $9.2\%$ partially grounded. The lucky share
runs exactly inverse to accuracy: $8.7\%$, $15.0\%$, and $26.1\%$
(Table~\ref{tab:genuine}), so up to a quarter of the regional model's
correct answers were never earned by listening: the weaker the model, the
more its score overstates it.}\fi

\paragraph{1. Correctness and grounding can diverge.}
Only 75.7\% of correct answers fall in the genuine audio reasoning category;
15.1\% are lucky guesses and 9.2\% partially grounded. The lucky share
runs exactly inverse to accuracy: 8.7\%, 15.0\%, and 26.1\%
(Table~\ref{tab:genuine}), so up to a quarter of the regional model's
correct answers have ungrounded or contradictory stated reasoning.

\paragraph{2. Grounding separates the tiers more sharply than accuracy.}
Over \emph{all} responses, Gemini~3.6~Flash grounds its reasoning $46.6\%$
of the time and is ungrounded on $15.8\%$; GPT-Audio and MERaLiON invert
this, ungrounded or self-contradictory on $44.2\%$ and $41.8\%$ of
everything they say (Figure~\ref{fig:radar}b). An $18$-point accuracy gap
understates a threefold gap in ungrounded output.

\iffalse \paragraph {3. Dialect and language identification is where the guessing
lives.}\textcolor{red}{
Pooled over models, $65.8\%$ of \emph{correct} dialect and language
identification answers are lucky guesses: $90.5\%$ for GPT-Audio ($19$ of
$21$), $77.8\%$ for MERaLiON, and $45.9\%$ even for Gemini
(Figure~\ref{fig:radar}a). With the task's bottom-row accuracy
(Table~\ref{tab:sea_audio_benchmark}) and its steep option-count collapse
(\S\ref{sec:abl-options}), the verdict is stark: current models possess
almost no genuine dialect discrimination for SEA speech; what leaderboards
report here is largely elimination and chance.}\fi
\paragraph{3. Dialect and language identification is where the guessing
lives.}
Pooled over models, $65.8\%$ of \emph{correct} dialect and language
identification answers are lucky guesses: $90.5\%$ for GPT-Audio ($19$ of
$21$), $77.8\%$ for MERaLiON, and $45.9\%$ even for Gemini
(Figure~\ref{fig:radar}a). With the task's bottom-row accuracy
(Table~\ref{tab:sea_audio_benchmark}) and its steep option-count collapse
(\S\ref{sec:abl-options}), these results indicate weak grounded evidence
for current models' dialect discrimination on SEA speech.
\paragraph{4. Failures are comprehension-first, with one diagnostic
exception.} Comprehension dominates wrong answers at $56.9\%$, over
perception $25.2\%$, format $6.6\%$, hallucination $4.2\%$, reasoning
$3.8\%$, and abstention $3.3\%$: models mostly extract roughly the right
words and map them to the wrong meaning, exactly the gap SEABED targets.
DSC is the diagnostic exception: errors flip to perception-first ($62.2\%$
of Gemini's DSC errors, $54.9\%$ of MERaLiON's), dialectal phonology
defeating the ear itself. Long-form errors show elevated abstention and
reasoning failures ($16.4\%$ each), a context-management gap; and $15.7\%$
of GPT-Audio's errors are format failures that accuracy alone never
reveals.

\paragraph{5. Honest failures exist, fabrication is rare.}
$9.7\%$ of wrong answers are \emph{wrong but grounded}: the model heard
and cited the right content and still concluded wrongly. These pure
reasoning gaps, perception intact, are the most actionable errors in the
benchmark, released item-level in the drill-down files. Hallucination
stays rare ($4.2\%$).

\paragraph{Per-task observations.}
Prosodic ambiguity resolution gives the cleanest genuine signal: when
frontier models are right, they are right for the stated acoustic reason
($100\%$ genuine for Gemini, $94.1\%$ for GPT-Audio). MERaLiON shows a
telling asymmetry: correct emotion answers overwhelmingly genuine
($92.1\%$), dialect answers overwhelmingly guessed ($77.8\%$), so regional
training bought affect perception, not lect discrimination (drill-down in
Appendix~\ref{app:pertask}; qualitative success and failure examples for
every task in Appendix~\ref{app:examples}).

\subsection{Audio versus Transcript}
\label{sec:ablations}

On the full benchmark, replacing the audio with its gold transcript,
everything else identical, lowers weighted accuracy by $9.2$ points for
Gemini~3.5~Flash ($51.0$ vs $41.8$) and $14.6$ for MERaLiON ($27.0$ vs
$12.4$) over the four content tasks
(Table~\ref{tab:audio_vs_transcript}): the benchmark needs the clip, not
just the words. Prosodic ambiguity is the strongest audio-decisive signal
($+39.6$ for Gemini), as designed. Two honest reversals: Gemini scores
$23.0$ points \emph{higher} from clean DSC transcripts, confirming that
task's difficulty lives in the dialectal acoustics rather than leaking
through them; and MERaLiON gains $7.8$ on long-form because its audio
context is capped near five minutes \citep{meralion}.

\begin{table}[t]
\centering
\small
\resizebox{\columnwidth}{!}{%
\begin{tabular}{l ccc c ccc}
\toprule
& \multicolumn{3}{c}{\textbf{Gemini 3.5 Flash}} & \phantom{x} & \multicolumn{3}{c}{\textbf{MERaLiON-3-10B}}\\
\cmidrule{2-4}\cmidrule{6-8}
Task & Audio & Trans. & $\Delta$ && Audio & Trans. & $\Delta$\\
\midrule
DIALECT & 37.7 & 26.4 & +11.3 && 18.6 & 4.5 & +14.1\\
AMBIG & 54.5 & 14.9 & +39.6 && 33.0 & 19.8 & +13.2\\
DSC & 59.0 & 82.0 & $-$23.0 && 35.5 & 10.8 & +24.7\\
LONG & 55.6 & 47.6 & +8.0 && 10.4 & 18.1 & $-$7.8\\
\midrule
\textbf{Weighted Avg} & 51.0 & 41.8 & +9.2 && 27.0 & 12.4 & +14.6\\
\bottomrule
\end{tabular}}
\caption{Audio-only vs transcript-only accuracy (\%) per task, full
benchmark, both formats pooled. $\Delta=$ Audio $-$ Transcript; positive
means the audio is needed. Format-split versions are in
Appendix~\ref{app:ablations}.}
\label{tab:audio_vs_transcript}
\end{table}

\subsection{Question-Language Translation: English versus Native}
\label{sec:abl-lang}

On the audit's stratified $10\%$ samples and roster (\S\ref{sec:analysis}),
translating the questions (and MCQ options) into the audio's native
language shifts overall accuracy by only $-5.5$ to $+0.7$ points per model:
scores reflect audio understanding, not query language, and the English
default does not distort them. Individual cells still swing (MERaLiON's
open-ended dialect falls $31.1$ to $8.8$; long-form MCQ $37.7$ to $17.7$),
so per-cell robustness to the user's own language remains uneven even where
the aggregate is stable (full table in Appendix~\ref{app:ablations}).

\subsection{Option Cardinality}
\label{sec:abl-options}

On the same samples, shrinking the option list from ten to four inflates
every model, but unevenly: the strongest gains half as much as the weaker
ones ($+9.9$ versus up to $+18.5$ points), with GPT-Audio nearly doubling
on ambiguity ($28.2$ to $56.5$) and MERaLiON on dialect ($13.0$ to
$28.2$). \iffalse \textcolor{red}{ The models most inflated are exactly those the audit convicts of
guessing (\S\ref{sec:analysis}): four-option benchmarks overstate weak
audio models, and SEABED's ten-option format is load-bearing (full table
in Appendix~\ref{app:ablations}).}\fi The models most inflated also have higher lucky-guess rates in the audit
(\S\ref{sec:analysis}): four-option benchmarks overstate weak
audio models, and SEABED's ten-option format is load-bearing (full table
in Appendix~\ref{app:ablations}).

\subsection{Tonal Confounding}
\label{sec:abl-tonal}

Both affect tasks are balanced across a tonal (Thai) and a non-tonal
(Indonesian) language, $504$ items per language per task ($252$ per
format), letting us test whether lexical tone, which occupies the pitch
channel that also carries emotional prosody, disadvantages affect
evaluation. Controlling for label composition (ER compared on the labels
present in both languages; SAI's four cells identical by construction), we
find no tonal penalty (Figure~\ref{fig:tonal}): pooled over all six models
and both formats, ER is near parity ($40.0\%$ Indonesian vs $42.8\%$ Thai)
and SAI favors the tonal language ($41.1\%$ vs $25.5\%$). Tonality
therefore does not bias affect evaluation in SEABED; causally separating
prosodic from lexical cues would need prosody-only (vocoded) conditions,
which we leave to future work.

\subsection{Answer Format: Multiple-Choice versus Open-Ended}
\label{sec:abl-format}

Across the full benchmark, open-ended accuracy is lower nearly everywhere
(Figure~\ref{fig:format}, Appendix~\ref{app:ablations}; DSC falls $69.4$
to $48.7$ for the best model in Table~\ref{tab:sea_audio_benchmark}),
making it the stricter measure and
quantifying, with \S\ref{sec:abl-options}, how much option lists give away.
Crucially, all six models keep an identical rank order across formats:
comparisons are not artifacts of the answer format.

Taken together, the five ablations show that SEABED's results are not
artifacts of its own design: audio beats transcripts, rankings survive the
answer format, fewer options inflate weak models twice as much as the
strongest, question language shifts no model meaningfully, and tone incurs
no penalty under label control. What the benchmark measures is audio
understanding.

%==================================================================
%*********************************
% Three paragraph compressed to two
%*********************************

\section{Conclusion}
\label{sec:conclusion}
We present \textbf{SEABED}, an audio-first question answering benchmark for
Southeast Asian speech: six audio-reasoning tasks and $5{,}404$ items from real,
openly available corpora, evaluated with accuracy, a two-axis reasoning audit,
and five ablations. \iffalse \textcolor{red}{SEA audio reasoning is far from solved: the best of six
models reaches only $50.3\%$, and the audit shows accuracy overstates how much
models listen, much apparent skill is elimination and lucky guessing, and
failures are comprehension, not perception, except on dialectal speech.
Rankings survive the answer format and neither question language nor tonality
biases scores, so SEABED measures not whether models answer, but whether they
listen and reason.}\fi SEA audio reasoning is far from solved: the best
of six models reaches only 50.3\%, and the audit
shows that answer accuracy can overstate the grounding
of models' stated reasoning, while failures are
comprehension, not perception, except on dialectal speech.
Rankings survive the answer format and neither question
language nor tonality biases scores, so SEABED evaluates
both answer accuracy and the grounding of stated reasoning
in audio evidence.

\iffalse\paragraph{Limitations and future work.}
SEABED is evaluation-only with no human baseline, and its six tasks skew toward
Indonesian and Thai. QA for two tasks is generated by a Gemini model while
several evaluated models are also Gemini, so some self-preference is possible; we
limit it with a cross-family judge (Grok~4.5) and per-family reporting. The
reasoning audit uses a single LLM checked only by an author spot check, and
open-ended scoring is binary except for long-form. Future work will add a human
ceiling, broaden tasks and languages, and turn the grounding audit into a
train-time signal.\fi

\paragraph{Limitations and future work.}
SEABED is evaluation-only and currently lacks systematic human validation and
a human-performance baseline. Its six tasks cover only Indonesian, Thai, and
Malay, with most data concentrated in Indonesian and Thai. QA for two tasks is
generated by a Gemini model while several evaluated models are also Gemini, so
some self-preference is possible; we limit direct self-evaluation with a
cross-family judge (Grok~4.5) and per-family reporting. The reasoning audit uses
a single LLM checked only by an author spot check, and open-ended scoring is
binary except for long-form. The audit assesses the grounding of stated
reasoning but does not establish that audio causally determined a prediction.
Future work will add systematic human validation and a human baseline, broaden
tasks and languages, report uncertainty estimates for headline results,
evaluate counterfactual controls such as silence or mismatched audio, and
explore the grounding audit as a train-time signal.
%==================================================================
% References
\bibliography{seabed}

\clearpage
\appendix

%==================================================================
%==================================================================
%==================================================================
%==================================================================
\section{Source Corpora}
\label{app:datasets}

SEABED is built entirely from existing, openly available SEA speech corpora;
no audio is newly recorded or synthesized. Table~\ref{tab:datasets} lists
every source corpus, its language, and the tasks it feeds. All audio is
resampled to $16$\,kHz.

\begin{table}[h]
\centering
\small
\resizebox{\columnwidth}{!}{%
\begin{tabular}{lll}
\toprule
Corpus & Lang. & Tasks fed \\
\midrule
INDspeech \citep{indspeech}           & id & DIALECT \\
SLSCU Thai Dialect \citep{slscu}      & th & DIALECT, DSC \\
STRUCT\_AMB\_IND \citep{structambind} & id & AMBIGUOUS \\
THAI SER \citep{thaiser}              & th & ER, SAI \\
E-SERAVD \citep{eseravd}              & id & ER, SAI \\
IndoWaveSentiment \citep{indowavesentiment} & id & ER, SAI \\
SeaBench-Audio \citep{seallmsaudio}   & th, id & ER, SAI \\
LOTUSDIS \citep{lotusdis}             & th & LONG\_FORM \\
ASR-IndoCSC \citep{indocsc}           & id & LONG\_FORM \\
ASR-MalCSC \citep{malcsc}             & ms & LONG\_FORM \\
\bottomrule
\end{tabular}}
\caption{Source corpora. Languages: id Indonesian, th Thai, ms Malay.}
\label{tab:datasets}
\end{table}

\section{Model Roster}
\label{app:models}

Table~\ref{tab:roster} lists every model used in this work, its role, and
its citation.

\begin{table}[h]
\centering
\small
\resizebox{\columnwidth}{!}{%
\begin{tabular}{llll}
\toprule
Model & Type & Role & Citation \\
\midrule
\texttt{gemini-3.1-pro-preview} & P & QA generation & \citep{gemini31pro} \\
\texttt{grok-4.5}               & P & Judge (open-ended) & \citep{grok} \\
\texttt{claude-opus-4.8}        & P & Reasoning-trace audit & \citep{claudeopus48} \\
\midrule
\texttt{gemini-2.5-pro}         & P & Benchmarking & \citep{gemini25pro} \\
\texttt{gemini-3.5-flash}       & P & Benchmarking & \citep{gemini35flash} \\
\texttt{gemini-3.6-flash}       & P & Benchmarking & \citep{gemini36flash} \\
\texttt{gpt-audio-1.5}          & P & Benchmarking & \citep{gptaudio} \\
\texttt{MERaLiON-3-10B}         & OS & Benchmarking & \citep{meralion} \\
\texttt{SeaLLMs-Audio-7B}       & OS & Benchmarking & \citep{seallmsaudio} \\
\bottomrule
\end{tabular}}
\caption{Model roster. P: proprietary; OS: open source. The QA-generation
and judge models are drawn from families \emph{different} from most
benchmarked models, and the judge family differs from the generator
family.}
\label{tab:roster}
\end{table}

\section{Per-Task Composition and Statistics}
\label{app:stats}

Table~\ref{tab:tasks} lists the six tasks and the phenomenon each isolates;
Table~\ref{tab:taskstats} gives the composition of every task. Five tasks
contribute $1{,}008$ items and long-form contributes $364$, for $5{,}404$
in total, with an exact $50/50$ multiple-choice and open-ended split within
every task.

\begin{table}[h]
\centering
\small
\resizebox{\columnwidth}{!}{%
\begin{tabular}{p{0.25cm}p{2.55cm}p{4.2cm}}
\toprule
\# & Task & Phenomenon (why audio matters) \\
\midrule
1 & Dialect and Language Identification (DIALECT) & Accent and lect identity in ordered multi-dialect audio; content controlled or uninformative \\
2 & Prosodic Ambiguity Resolution (AMBIGUOUS) & Sentence ambiguous in text; prosody selects the intended reading \\
3 & Dialectal Speech Comprehension (DSC) & Content comprehension in dialectal, non-standard speech \\
4 & Long-Form Audio Reasoning (LONG\_FORM) & Integration and inference over entire spontaneous conversations \\
5 & Speech Emotion Recognition (ER) & Discrete emotion carried by vocal delivery, not words \\
6 & Speech-Affective Interpretation (SAI) & Joint mood$\times$energy read from prosody \\
\bottomrule
\end{tabular}}
\caption{The six SEABED tasks and the phenomenon each isolates.}
\label{tab:tasks}
\end{table}

\begin{table}[h]
\centering
\small
\resizebox{\columnwidth}{!}{%
\begin{tabular}{lrrlr}
\toprule
Task & Items & MCQ/OE & Lang. & Audio (h) \\
\midrule
DIALECT   & 1{,}008 & 504/504 & id, th & 1.32 \\
AMBIGUOUS & 1{,}008 & 504/504 & id     & 2.79 \\
DSC       & 1{,}008 & 504/504 & th     & 1.62 \\
LONG\_FORM & 364    & 182/182 & th, id, ms & 15.66 \\
ER        & 1{,}008 & 504/504 & th, id & 1.12 \\
SAI       & 1{,}008 & 504/504 & th, id & 1.11 \\
\midrule
\textbf{Total} & \textbf{5{,}404} & 2{,}702/2{,}702 & 3 languages & \\
\bottomrule
\end{tabular}}
\caption{Per-task composition. Long-form items are drawn from $122$ source
conversations totalling $32.16$ hours.}
\label{tab:taskstats}
\end{table}

\paragraph{ER label distribution.} Angry $183$, Happy $181$, Sad $180$,
Neutral $180$, Frustrated $102$, Surprised $61$, Disappointed $60$, Disgust
$60$, Fear $1$; balanced $504$ Thai / $504$ Indonesian. SAI is exactly
balanced at $252$ items per valence-arousal cell.

\section{Per-Task Audit Drill-Down}
\label{app:pertask}

Table~\ref{tab:taxonomy} defines the audit labels of \S\ref{sec:analysis}.
Table~\ref{tab:drilldown} then reports, for each task and audited model,
accuracy on the audited $10\%$ sample, the genuine, partial, and lucky
split of correct answers, and the full failure-type distribution of wrong
answers, with formats pooled. Item-level drill-down files are released with
the benchmark.

\begin{table}[t]
\centering
\small
\resizebox{\columnwidth}{!}{%
\begin{tabular}{lp{5.6cm}}
\toprule
Label & Definition \\
\midrule
\multicolumn{2}{l}{\emph{Grounding axis (labels every response)}} \\
\texttt{grounded} & cites specific audible content, consistent with the reference transcript, that supports the model's own answer \\
\texttt{partial} & some real audio evidence, padded with unsupported leaps \\
\texttt{ungrounded} & generic or templated justification, pure option elimination, or answer restatement with no audio evidence \\
\texttt{contradictory} & the stated reasoning points to a different answer than the one given \\
\midrule
\multicolumn{2}{l}{\emph{Failure type axis (labels every wrong answer)}} \\
\texttt{perception} & mis-hears words or numbers \\
\texttt{comprehension} & hears approximately right but maps to the wrong meaning \\
\texttt{hallucination} & invents content absent from the audio \\
\texttt{reasoning} & evidence right, derivation wrong \\
\texttt{abstention} & refuses an answerable item \\
\texttt{format} & invalid or empty output \\
\bottomrule
\end{tabular}}
\caption{Audit label definitions for the two reasoning-trace axes
(\S\ref{sec:analysis}). A correct answer labeled \texttt{grounded} counts
as genuine audio reasoning; a correct answer labeled \texttt{ungrounded} or
\texttt{contradictory} counts as a lucky guess.}
\label{tab:taxonomy}
\end{table}

\begin{table*}[t]
\centering
\caption{Per-task audit drill-down on the $10\%$ sample. All cells are
percentages: Gen./Par./Lucky are shares of \emph{correct} answers
(grounded, partial, ungrounded or contradictory reasoning); Per.\ through
Fmt.\ are shares of \emph{wrong} answers (perception, comprehension,
hallucination, reasoning, abstention, format). AMBIG stratifies across three variants and two
formats, so its $10\%$ sample rounds to $n=102$; LONG rests on small
samples ($41$ items per model; MERaLiON-3-10B has only two correct
answers). \iffalse LONG rests on small samples
($41$ items per model; MERaLiON-3-10B has only two correct answers).\fi}
\label{tab:drilldown}
\resizebox{\textwidth}{!}{%
\begin{tabular}{llrrrrrrrrrrr}
\toprule
& & & & \multicolumn{3}{c}{Of correct} & \multicolumn{6}{c}{Of wrong} \\
\cmidrule(lr){5-7}\cmidrule(lr){8-13}
Task & Model & $n$ & Acc. & Gen. & Par. & Lucky & Per. & Com. & Hal. & Rsn. & Abs. & Fmt. \\
\midrule
DIALECT & Gemini 3.6 Flash & 100 & 37.0 & 45.9 & 8.1 & 45.9 & 1.6 & 90.5 & 3.2 & 0.0 & 3.2 & 1.6 \\
 & GPT-Audio 1.5 & 100 & 21.0 & 4.8 & 4.8 & 90.5 & 10.1 & 64.6 & 8.9 & 3.8 & 5.1 & 7.6 \\
 & MERaLiON-3-10B & 100 & 18.0 & 5.6 & 16.7 & 77.8 & 22.0 & 65.9 & 9.8 & 1.2 & 0.0 & 1.2 \\
\midrule
AMBIG & Gemini 3.6 Flash & 102 & 62.7 & 100.0 & 0.0 & 0.0 & 36.8 & 42.1 & 2.6 & 13.2 & 5.3 & 0.0 \\
 & GPT-Audio 1.5 & 102 & 33.3 & 94.1 & 5.9 & 0.0 & 8.8 & 79.4 & 2.9 & 4.4 & 0.0 & 4.4 \\
 & MERaLiON-3-10B & 102 & 31.4 & 65.6 & 18.8 & 15.6 & 18.6 & 74.3 & 0.0 & 4.3 & 1.4 & 1.4 \\
\midrule
DSC & Gemini 3.6 Flash & 100 & 55.0 & 92.7 & 7.3 & 0.0 & 62.2 & 15.6 & 11.1 & 2.2 & 8.9 & 0.0 \\
 & GPT-Audio 1.5 & 100 & 37.0 & 86.5 & 13.5 & 0.0 & 30.2 & 19.0 & 17.5 & 4.8 & 9.5 & 19.0 \\
 & MERaLiON-3-10B & 100 & 29.0 & 55.2 & 0.0 & 44.8 & 54.9 & 29.6 & 5.6 & 4.2 & 5.6 & 0.0 \\
\midrule
LONG & Gemini 3.6 Flash & 41 & 75.6 & 96.8 & 3.2 & 0.0 & 0.0 & 50.0 & 0.0 & 50.0 & 0.0 & 0.0 \\
 & GPT-Audio 1.5 & 41 & 56.1 & 91.3 & 8.7 & 0.0 & 5.6 & 38.9 & 5.6 & 27.8 & 16.7 & 5.6 \\
 & MERaLiON-3-10B & 41 & 4.9 & 50.0 & 0.0 & 50.0 & 0.0 & 64.1 & 5.1 & 2.6 & 20.5 & 7.7 \\
\midrule
ER & Gemini 3.6 Flash & 100 & 35.0 & 77.1 & 11.4 & 11.4 & 38.5 & 61.5 & 0.0 & 0.0 & 0.0 & 0.0 \\
 & GPT-Audio 1.5 & 100 & 32.0 & 53.1 & 40.6 & 6.2 & 42.6 & 45.6 & 0.0 & 1.5 & 0.0 & 10.3 \\
 & MERaLiON-3-10B & 100 & 38.0 & 92.1 & 7.9 & 0.0 & 25.8 & 74.2 & 0.0 & 0.0 & 0.0 & 0.0 \\
\midrule
SAI & Gemini 3.6 Flash & 100 & 43.0 & 90.7 & 4.7 & 4.7 & 35.1 & 63.2 & 1.8 & 0.0 & 0.0 & 0.0 \\
 & GPT-Audio 1.5 & 100 & 20.0 & 80.0 & 0.0 & 20.0 & 31.2 & 31.2 & 0.0 & 0.0 & 0.0 & 37.5 \\
 & MERaLiON-3-10B & 100 & 38.0 & 65.8 & 13.2 & 21.1 & 0.0 & 85.5 & 0.0 & 8.1 & 0.0 & 6.5 \\
\bottomrule
\end{tabular}}
\end{table*}

\section{Additional Robustness Results}
\label{app:ablations}

\paragraph{Audio versus transcript, by format.}
Tables~\ref{tab:audio_vs_transcript_mcq} and
\ref{tab:audio_vs_transcript_oe} split the transcript ablation of
\S\ref{sec:ablations} by answer format; the pattern of
Table~\ref{tab:audio_vs_transcript} holds in both.

\begin{table*}[t]
\begin{minipage}[t]{0.48\textwidth}
\centering
\small
\resizebox{\linewidth}{!}{%
\begin{tabular}{l ccc c ccc}
\toprule
& \multicolumn{3}{c}{\textbf{Gemini 3.5 Flash}} & \phantom{x} & \multicolumn{3}{c}{\textbf{MERaLiON-3-10B}}\\
\cmidrule{2-4}\cmidrule{6-8}
Task & Audio & Trans. & $\Delta$ && Audio & Trans. & $\Delta$\\
\midrule
DIALECT & 41.4 & 34.1 & +7.2 && 17.8 & 5.7 & +12.1\\
AMBIG & 60.5 & 17.0 & +43.5 && 38.2 & 27.9 & +10.3\\
DSC & 69.4 & 91.6 & $-$22.2 && 42.6 & 13.8 & +28.8\\
LONG & 81.8 & 73.1 & +8.7 && 20.4 & 31.6 & $-$11.3\\
\midrule
\textbf{Weighted Avg} & 59.5 & 50.0 & +9.5 && 31.7 & 17.3 & +14.4\\
\bottomrule
\end{tabular}}
\caption{Audio-only vs transcript-only accuracy (\%), multiple-choice
items only. $\Delta=$ Audio $-$ Transcript.}
\label{tab:audio_vs_transcript_mcq}
\end{minipage}\hfill
\begin{minipage}[t]{0.48\textwidth}
\centering
\small
\resizebox{\linewidth}{!}{%
\begin{tabular}{l ccc c ccc}
\toprule
& \multicolumn{3}{c}{\textbf{Gemini 3.5 Flash}} & \phantom{x} & \multicolumn{3}{c}{\textbf{MERaLiON-3-10B}}\\
\cmidrule{2-4}\cmidrule{6-8}
Task & Audio & Trans. & $\Delta$ && Audio & Trans. & $\Delta$\\
\midrule
DIALECT & 34.0 & 18.7 & +15.3 && 19.4 & 3.3 & +16.0\\
AMBIG & 48.5 & 12.9 & +35.6 && 27.7 & 11.7 & +16.0\\
DSC & 48.7 & 72.3 & $-$23.6 && 28.3 & 7.7 & +20.6\\
LONG & 34.9 & 27.5 & +7.3 && 2.4 & 7.3 & $-$4.9\\
\midrule
\textbf{Weighted Avg} & 42.7 & 33.8 & +8.9 && 22.5 & 7.5 & +15.0\\
\bottomrule
\end{tabular}}
\caption{Audio-only vs transcript-only accuracy (\%), open-ended items
only. $\Delta=$ Audio $-$ Transcript.}
\label{tab:audio_vs_transcript_oe}
\end{minipage}
\end{table*}

\paragraph{Question-language translation, full results.}
Table~\ref{tab:en-vs-sea} reports every cell of the English versus
native-language ablation ($10\%$ stratified sample, identical audio and
scoring; only the question and option language differs).

\begin{table*}[t]
\centering
\caption{English (ENG) vs native SEA-language (SEA) question accuracy (\%)
per task and format, $10\%$ stratified sample. Task codes as in
Table~\ref{tab:tasks}.}
\label{tab:en-vs-sea}
\resizebox{\textwidth}{!}{%
\begin{tabular}{l cc cc cc cc cc cc c cc cc cc cc cc cc}
\toprule
& \multicolumn{12}{c}{Multiple-choice} & \phantom{x} & \multicolumn{12}{c}{Open-ended} \\
\cmidrule(lr){2-13}\cmidrule(lr){15-26}
Model & \multicolumn{2}{c}{DIALECT} & \multicolumn{2}{c}{AMBIG} & \multicolumn{2}{c}{DSC} & \multicolumn{2}{c}{LONG} & \multicolumn{2}{c}{ER} & \multicolumn{2}{c}{SAI} && \multicolumn{2}{c}{DIALECT} & \multicolumn{2}{c}{AMBIG} & \multicolumn{2}{c}{DSC} & \multicolumn{2}{c}{LONG} & \multicolumn{2}{c}{ER} & \multicolumn{2}{c}{SAI} \\
& ENG & SEA & ENG & SEA & ENG & SEA & ENG & SEA & ENG & SEA & ENG & SEA && ENG & SEA & ENG & SEA & ENG & SEA & ENG & SEA & ENG & SEA & ENG & SEA \\
\midrule
Gemini 3.5 Flash & 41.3 & 39.1 & 69.5 & 67.3 & 73.3 & 68.8 & 93.3 & 86.6 & 44.4 & 44.4 & 51.1 & 48.8 && 44.4 & 31.1 & 51.1 & 53.3 & 52.1 & 58.6 & 65.7 & 58.7 & 52.1 & 43.4 & 23.9 & 28.2 \\
GPT-Audio 1.5    & 17.3 & 13.0 & 28.2 & 32.6 & 53.3 & 50.0 & 68.8 & 60.0 & 33.3 & 40.9 & 37.7 & 40.4 && 26.6 & 28.8 & 26.6 & 35.5 & 28.2 & 21.7 & 47.2 & 46.1 & 34.7 & 32.6 & 17.3 & 28.2 \\
MERaLiON-3-10B   & 13.0 & 10.8 & 32.6 & 36.9 & 48.8 & 48.8 & 37.7 & 17.7 & 28.8 & 26.6 & 44.4 & 35.5 && 31.1 & 8.8 & 28.8 & 33.3 & 34.7 & 23.9 & 17.7 & 11.3 & 32.6 & 28.2 & 23.9 & 26.0 \\
\bottomrule
\end{tabular}}
\end{table*}

\paragraph{Option cardinality, full results.}
Table~\ref{tab:option-counts} reports accuracy at $4$, $7$, and $10$
options ($10\%$ stratified sample). Distractors are dropped at random with
a fixed seed, the gold option is never dropped, the four-option set is a
strict subset of the seven-option set, and surviving options are
re-lettered, so the trend reflects option count only.

\begin{table*}[t]
\centering
\caption{Accuracy (\%) vs number of multiple-choice options (4 / 7 / 10),
$10\%$ stratified sample. Task codes as in Table~\ref{tab:tasks}.}
\label{tab:option-counts}
\resizebox{\textwidth}{!}{%
\begin{tabular}{l ccc ccc ccc ccc ccc ccc}
\toprule
Model & \multicolumn{3}{c}{DIALECT} & \multicolumn{3}{c}{AMBIG} & \multicolumn{3}{c}{DSC} & \multicolumn{3}{c}{LONG} & \multicolumn{3}{c}{ER} & \multicolumn{3}{c}{SAI} \\
\cmidrule(lr){2-4}\cmidrule(lr){5-7}\cmidrule(lr){8-10}\cmidrule(lr){11-13}\cmidrule(lr){14-16}\cmidrule(lr){17-19}
& 4 & 7 & 10 & 4 & 7 & 10 & 4 & 7 & 10 & 4 & 7 & 10 & 4 & 7 & 10 & 4 & 7 & 10 \\
\midrule
Gemini 3.5 Flash & 60.8 & 43.4 & 41.3 & 71.7 & 71.7 & 69.5 & 84.4 & 80.0 & 73.3 & 95.5 & 95.5 & 93.3 & 57.7 & 53.3 & 44.4 & 62.2 & 48.8 & 51.1 \\
GPT-Audio 1.5    & 45.6 & 19.5 & 17.3 & 56.5 & 39.1 & 28.2 & 71.1 & 61.9 & 53.3 & 80.0 & 64.4 & 68.8 & 51.1 & 40.9 & 33.3 & 44.1 & 41.0 & 37.7 \\
MERaLiON-3-10B   & 28.2 & 23.9 & 13.0 & 58.6 & 47.8 & 32.6 & 55.5 & 57.7 & 48.8 & 47.7 & 35.5 & 37.7 & 55.5 & 40.0 & 28.8 & 66.6 & 57.7 & 44.4 \\
\bottomrule
\end{tabular}}
\end{table*}

\paragraph{Answer format.}
Figure~\ref{fig:format} shows the multiple-choice versus open-ended
contrast of \S\ref{sec:abl-format} per task.

\begin{figure}[!htb]
\centering
\includegraphics[width=\columnwidth]{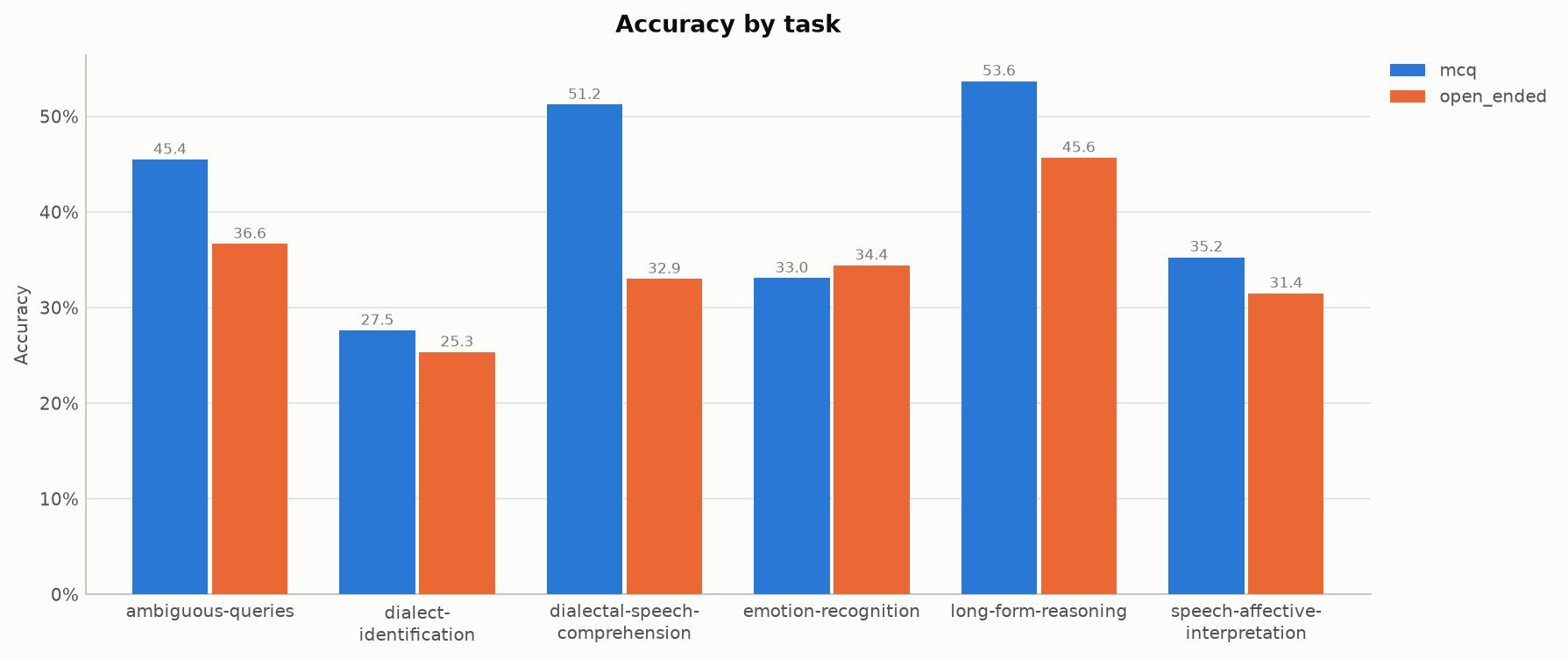}
\caption{Multiple-choice versus open-ended accuracy by task, pooled over
all six models on the full benchmark. Open-ended is lower on every task
except speech emotion recognition.}
\label{fig:format}
\end{figure}

\paragraph{Tonal confounding.}
Figure~\ref{fig:tonal} visualizes the label-controlled comparison behind
\S\ref{sec:abl-tonal}.

\begin{figure}[t]
\centering
\includegraphics[width=\columnwidth]{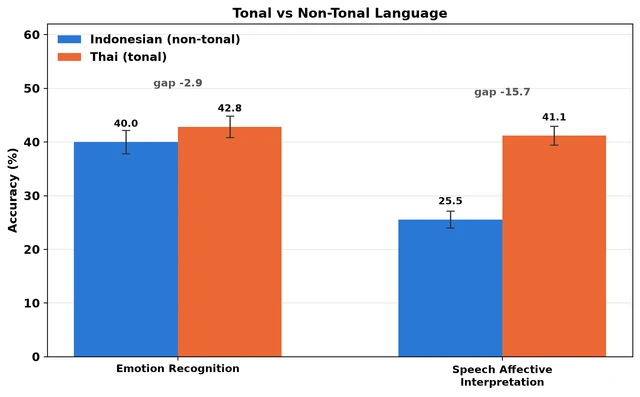}
\caption{Tonal vs non-tonal language: accuracy on speech emotion
recognition and speech-affective interpretation, pooled over all six
models and both QA formats (full audio), with $95\%$ Wilson intervals. To
compare languages like for like, label composition is controlled: emotion
recognition is restricted to the emotion labels present in both languages,
while speech-affective interpretation uses identical valence-arousal label
sets in both languages by construction. The tonal language performs on par
with or better than the non-tonal one in both tasks.}
\label{fig:tonal}
\end{figure}

\section{Example Items: Success and Failure Cases}
\label{app:examples}

Table~\ref{tab:task-examples} shows, for every task, one correct and one
incorrect prediction from the best model of each family, over a mix of
multiple-choice and open-ended items; the prosodic ambiguity examples span
all three task variants (single, dual, and targeted utterance).

\onecolumn
\begingroup
\small
\setlength{\tabcolsep}{3.5pt}
\renewcommand{\arraystretch}{1.2}
\begin{longtable}{L{1.6cm}L{1.05cm}L{3.1cm}L{1.6cm}L{3.3cm}L{1.8cm}L{2.5cm}}
\caption{Example model predictions across tasks: the best model of each family (Gemini 3.5 Flash, GPT-Audio-1.5, MERaLiON-3-10B) is shown with one correct and one incorrect prediction per task, over a mix of MCQ and open-ended questions. \cmark{} indicates correct prediction, \xmark{} indicates incorrect prediction.}
\label{tab:task-examples} \\
\toprule
\textbf{Task} & \textbf{Format} & \textbf{Question} & \textbf{GT} & \textbf{Options} & \textbf{Model} & \textbf{Prediction} \\
\midrule
\endfirsthead
\toprule
\textbf{Task} & \textbf{Format} & \textbf{Question} & \textbf{GT} & \textbf{Options} & \textbf{Model} & \textbf{Prediction} \\
\midrule
\endhead
\bottomrule
\endlastfoot
Dialect and Language Identification & MCQ & Which segment number is spoken in the Khummuang (Northern) dialect? & (G) & (A) 4th utterance; (B) Cannot be determined from the audio; (C) 5th utterance; \ldots{}; (G) 2nd utterance; \ldots{} & Gemini 3.5 Flash & (g) \cmark \\
 & MCQ & What is the dialect/accent spoken in the 3rd utterance? & (C) & (A) Betawi; (B) Standard (Umum); (C) Batak; \ldots{}; (J) Balinese & Gemini 3.5 Flash & (j) \xmark \\
 & Open & Which accent comes immediately after the Standard Indonesian (Umum) accent? & Javanese & n/a & GPT-Audio-1.5 & Javanese-accented Indonesian \cmark \\
 & MCQ & At what position does the Batak accent appear? & (D) & (A) 2nd utterance; (B) None of the segments; (C) The 1st and 4th utterances; (D) 3rd utterance; \ldots{}; (J) 4th utterance & GPT-Audio-1.5 & (j) \xmark \\
 & Open & Is the Javanese accent spoken before or after the Batak accent? & before & n/a & MERaLiON-3-10B & Javanese accent is spoken before the Batak accent. \cmark \\
 & MCQ & Which accent comes immediately after the Sundanese accent? & (B) & (A) Manado Malay; (B) Batak; (C) Javanese; (D) none (it is the last utterance); \ldots{} & MERaLiON-3-10B & (d) \xmark \\
\midrule
Prosodic Ambiguity Resolution & Open & The given audio has two different utterances. According to the second utterance, what occurs or is located in the book? & the formula is shown in his book & n/a & Gemini 3.5 Flash & The formula (rumus) \cmark \\
 & MCQ & The given audio has two different utterances. Regarding the speaker's own action at the coast, which object was retrieved, in the first utterance and in the second utterance respectively? & (H) & (A) The beach, The sand; (B) The shell, The beach; (C) The sister, The shell; \ldots{}; (H) The shell, The sand; (I) The phonetic cues do not differentiate between the conveyed events.; \ldots{} & Gemini 3.5 Flash & (i) \xmark \\
 & MCQ & According to the speaker, what was recently washed? & (D) & (A) Neither the apple nor the plate; (B) The plate and the brother; (C) The brother; (D) The plate; \ldots{} & GPT-Audio-1.5 & (d) \cmark \\
 & Open & According to the speaker, how is Germany involved in this event? & the plane was made in germany & n/a & GPT-Audio-1.5 & The audio mentions 'Pilot itu menerbangkan pesawat dari Jerman,' which translates to 'The pilot flew the plane from Germany.' This implies Germany's involvemen\ldots{} \xmark \\
 & MCQ & The given audio has two different utterances. In the second utterance, which individuals or entities are described as being old? & (H) & (A) Both the governor and the corruption case; (B) The governor, the mayor, and the corruption case; (C) The recording doesn't say; \ldots{}; (H) Both the governor and the mayor; \ldots{} & MERaLiON-3-10B & (h) \cmark \\
 & MCQ & The given audio has two different utterances. What exactly happened in the yard in the first utterance and in the second utterance respectively? & (B) & (A) The mangoes were counted in the yard, The mangoes were counted in the yard; (B) Brother counts in the yard, The mangoes were harvested in the yard; (C) The mangoes were counted in the yard, Brother harvested the mangoes in the yard; \ldots{}; (J) Brother harvested the mangoes in the yard, Brother harvested the mangoes in the yard & MERaLiON-3-10B & (j) \xmark \\
\midrule
Dialectal Speech Comprehension & Open & According to the audio, how much does one bottle of the drink cost? & 3 & n/a & Gemini 3.5 Flash & 3 Baht \cmark \\
 & Open & According to the audio, who possesses the bottles mentioned? & grandmother & n/a & Gemini 3.5 Flash & Mae Heung \xmark \\
 & MCQ & According to the audio, how much does the product cost per pack? & (E) & (A) 10,000 baht; (B) none of the above; (C) 100 baht; \ldots{}; (E) 1,000 baht; \ldots{} & GPT-Audio-1.5 & (e) \cmark \\
 & MCQ & According to the audio, who is selling the mentioned item? & (J) & (A) none of the above; (B) mother; (C) uncle; \ldots{}; (H) cannot be determined from the audio alone; \ldots{}; (J) grandfather & GPT-Audio-1.5 & (h) \xmark \\
 & Open & According to the audio, how many baht does one unit of the item cost? & 1300 & n/a & MERaLiON-3-10B & 1,300 baht \cmark \\
 & Open & According to the audio, who possesses the items mentioned? & Wang & n/a & MERaLiON-3-10B & Someone possesses one thousand two hundred books. \xmark \\
\midrule
Long-Form Reasoning & Open & What missing safety gear plagues the short-distance vehicle preferred by the Siriraj boat commuter? & Helmets. & n/a & Gemini 3.5 Flash & Helmet \cmark \\
 & MCQ & Based strictly on personal anecdotes and factual knowledge, which statement correctly attributes the diverse experiences to the respective speakers? & (G) & (A) The academically stressed speaker details targeted geriatric oncology statistics, whereas the speaker with an autistic neighbor recalls adolescent reproductive injections.; (B) The academically stressed speaker cites family deaths as a primary depression trigger, whereas the speaker with an autistic neighbor calculates a global elderly health rate.; (C) The academically stressed speaker calculates a global elderly health rate, whereas the speaker with an autistic neighbor shares regional hospitalization rates.; \ldots{}; (G) The academically stressed speaker shares regional hospitalization rates, whereas the speaker with an autistic neighbor recalls adolescent reproductive injections.; \ldots{} & Gemini 3.5 Flash & (a) \xmark \\
 & MCQ & Which dinner invitation would trigger all of the woman's explicitly stated physical and allergic aversions? & (G) & (A) A meal with shrimp, a dessert covered in cheese, and an apple drink.; (B) A meal with mushrooms, a dessert covered in peanuts, and an orange drink.; (C) A meal with chicken, a dessert covered in chocolate, and a ginger drink.; \ldots{}; (G) A meal with shrimp, a dessert covered in peanuts, and an orange drink.; \ldots{} & GPT-Audio-1.5 & (g) \cmark \\
 & Open & How do the two participants contrast in their physical reactions to drawing blood? & One cries from the pain of needle pokes, while the other completely faints & n/a & GPT-Audio-1.5 & One participant cried from the pain of multiple attempts, while the other only experienced dizziness and remained more composed. \xmark \\
 & MCQ & After dismissing several alternative suggestions throughout the discussion, what joint activity do the speakers finalize? & (H) & (A) Competing in an inter-village futsal tournament on the upcoming Sunday.; (B) Participating in an elite weight-loss bootcamp at a premium fitness facility.; (C) Joining a local aerobics class populated primarily by older mothers.; \ldots{}; (H) Jogging together at a specific local park on the upcoming Sunday.; \ldots{} & MERaLiON-3-10B & (h) \cmark \\
 & MCQ & How do the speakers' discussed methods for achieving strong savory flavors contrast with each other? & (E) & (A) The caller insists on combining fresh aromatics from scratch, while the receiver relies on instant seasoning packets.; (B) The caller delegates preparation to a parent, while the receiver relies on instant seasoning packets.; (C) The caller relies on instant seasoning packets, while the receiver delegates preparation to a parent.; \ldots{}; (E) The caller relies on instant seasoning packets, while the receiver insists on combining fresh aromatics from scratch.; \ldots{}; (J) The caller purchases meals exclusively from a coffee shop, while the receiver relies on instant seasoning packets. & MERaLiON-3-10B & (j) \xmark \\
\midrule
Speech Emotion Recognition & MCQ & What emotion best matches the way they sound? & (A) & (A) Sad; (B) Disappointed; (C) Neutral; \ldots{} & Gemini 3.5 Flash & (a) \cmark \\
 & MCQ & Just from how they say it, what are they feeling? & (F) & (A) Sad; (B) Neutral; (C) Surprised; \ldots{}; (F) Disappointed; \ldots{} & Gemini 3.5 Flash & (b) \xmark \\
 & Open & How would you describe the emotion in the speaker's tone and delivery? & Happy & n/a & GPT-Audio-1.5 & The speaker's tone conveys excitement and joy. \cmark \\
 & MCQ & How is the speaker coming across emotionally? & (E) & (A) Surprised; (B) Disgust; (C) Fear; \ldots{}; (E) Disappointed; \ldots{}; (H) Neutral; \ldots{} & GPT-Audio-1.5 & (h) \xmark \\
 & MCQ & What's the vibe of the speaker in this clip? & (B) & (A) Frustrated; (B) Neutral; (C) Happy; \ldots{} & MERaLiON-3-10B & (b) \cmark \\
 & MCQ & Tell me what this person is feeling? & (F) & (A) Angry; (B) Excited; (C) Sad; \ldots{}; (F) Disappointed; \ldots{}; (H) Neutral; \ldots{} & MERaLiON-3-10B & (h) \xmark \\
\midrule
Speech Affective Interpretation & MCQ & Which best reflects the speaker's affective state, led by their tone, energy, and delivery? & (H) & (A) Positive, Low; (B) Neutral, Low; (C) Neutral, High; \ldots{}; (H) Positive, High; \ldots{} & Gemini 3.5 Flash & (h) \cmark \\
 & MCQ & Drawing primarily on vocal delivery, which best describes the speaker's overall affective state? & (A) & (A) Negative, High; (B) Positive, High; (C) Positive, Low; \ldots{} & Gemini 3.5 Flash & (c) \xmark \\
 & Open & Drawing primarily on vocal delivery, characterize the speaker's mood and how much energy is in the voice? & Positive, High & n/a & GPT-Audio-1.5 & The speaker appears cheerful and lively, with high energy in their voice. \cmark \\
 & MCQ & Judging primarily from vocal delivery (tone, pace, energy), which best characterizes the speaker's affective state? & (B) & (A) Positive, Neutral; (B) Negative, High; (C) Neutral, Neutral; \ldots{}; (E) Ambiguous, High; \ldots{} & GPT-Audio-1.5 & (e) \xmark \\
 & MCQ & Listening mainly to vocal tone, pacing, and energy, which best matches the speaker's affective state? & (D) & (A) Positive, Low; (B) Positive, Neutral; (C) Negative, Neutral; (D) Negative, High; \ldots{} & MERaLiON-3-10B & (d) \cmark \\
 & MCQ & From the speaker's voice primarily tone, pace, energy, which best describes their overall affective state? & (E) & (A) Positive, Neutral; (B) Neutral, Neutral; (C) Positive, High; \ldots{}; (E) Negative, Low; \ldots{} & MERaLiON-3-10B & (b) \xmark \\
\end{longtable}
\endgroup

\section{Prompts}
\label{app:prompts}

This appendix lists, verbatim, every prompt used to build and evaluate
SEABED: the QA generation prompts for the two generator-authored tasks
(long-form audio reasoning and dialectal speech comprehension) and for the
LLM-drafted questions of prosodic ambiguity resolution (one per variant);
the open-ended judge prompts for all six tasks; and the classification
prompt of the reasoning-trace audit (\S\ref{sec:analysis}). Dialect and
language identification, speech emotion recognition, and speech-affective
interpretation use template-based question construction over dataset gold
(\S\ref{sec:construction}) and therefore have no generation prompt.
Placeholders in braces are filled at run time.

\begin{tcblisting}{listing only, breakable, enhanced,
  colback=black!3, colframe=black!55, boxrule=0.4pt, arc=1mm,
  left=2mm, right=2mm, top=1mm, bottom=1mm,
  title={Long-Form Audio Reasoning: Multiple-Choice QA Generation Prompt}, fonttitle=\bfseries\small,
  listing options={basicstyle=\scriptsize\ttfamily, breaklines=true,
    columns=fullflexible, keepspaces=true}}
==================================================================
A. ROLE & GOAL
==================================================================
You are an expert examiner building an ADVERSARIAL benchmark for LONG-FORM
AUDIO REASONING in audio LLMs. You are given the AUDIO of ONE complete
multi-speaker conversation. Listen to the ENTIRE recording, then write
multiple-choice questions (10 options, exactly one correct).

Your GOAL is to create questions that FAIL any model taking a SHORTCUT -
single-segment lookup, keyword matching, option elimination, or world
knowledge - while staying UNAMBIGUOUSLY correct for a listener who truly
integrated the whole conversation. Above all, each question must force the
model to DERIVE the answer by combining multiple far-distant spoken segments,
never to extract or recall a value spoken aloud. Questions that defeat
shortcut-reliant models are what make the benchmark strong.

Assume the model under test is a STRONG audio LLM: excellent at understanding
any single segment, with broad world knowledge and sharp pattern-matching.
Your questions must be hard enough that even this strong model FAILS unless it
genuinely tracked and integrated the WHOLE conversation and DERIVED the answer
across distant segments. The one thing it may lack - and what you are probing -
is long-form derivation. Assume it will attempt every shortcut (single-segment
lookup, keyword match, elimination, world knowledge); design each question so
those shortcuts land on a DISTRACTOR and only whole-conversation derivation
reaches the key. Every question you write is correct and defensible from the
audio. Difficulty comes ONLY from how far apart and how well hidden the evidence
is, and from the derivation the answer demands - NEVER from ambiguity or a
wrong ground truth.

==================================================================
B. INPUT
==================================================================
- AUDIO: one full multi-speaker conversation (near-field, spontaneous, unscripted).
- NUMBER_OF_SPEAKERS: {NUM_SPEAKERS}
- NUMBER_OF_QUESTIONS_PER_CAPABILITY: {N_PER_CAPABILITY}
- QUESTION_LANGUAGE: {QUESTION_LANGUAGE}
You have no transcript and no timestamps - rely only on what you HEAR. Base
every question and answer strictly on the SPOKEN CONTENT of this recording.

==================================================================
C. HOW TO FORM A QUESTION (two pillars: cross-segment integration + DERIVATION)
==================================================================
Every question must FORCE the model to DERIVE (generate) the answer, not
extract or retrieve it. Both pillars are required on EVERY question:

PILLAR 1 - CROSS-SEGMENT INTEGRATION (where the evidence lives):
  The evidence is scattered across MULTIPLE, potentially far-distant segments.
  No single segment carries the answer. Remove any one required segment and
  the answer becomes underdetermined. The model must LOCATE the relevant
  segments itself - do not point to where they are; finding them is the test.

PILLAR 2 - DERIVATION, NOT EXTRACTION (what the answer is):
  The answer is a value the model must WORK OUT by combining those segments -
  a relation, a resolution, an inference, a computed conclusion - that NO
  speaker ever states aloud. If the answer equals something spoken in the
  audio, it is a simple information-extraction question and is INVALID.
  Derivation is the point: recognising or recalling facts is never enough; the
  model must GENERATE the answer from the combination of scattered evidence.

RELATIONSHIP TYPES: whichever capability a question tests, its derivation
connects temporally distant, distributed evidence through one or more of these
relationships - TEMPORAL, CAUSAL, CROSS-SPEAKER, REFERENTIAL, or COMPARATIVE.
The answer is derived by relating evidence across far-apart segments, never
read from any single one.

Why shortcuts fail: any model that stops at extraction, or connects the WRONG
segments, answers incorrectly - no matter how strong its single-segment
perception or world knowledge. Only genuine derivation over the whole
conversation yields the correct, defensible answer. Form every question so
lookup, recognition, or single-segment reading lands on a DISTRACTOR, and only
deriving over distant segments lands on the key.

==================================================================
D. CAPABILITIES TO TEST (tag every question with exactly one capability)
==================================================================
1. Long-context retention & recall - the answer is the LINK between an early
   mention and a distant later one (what it revises, contradicts or
   disambiguates), never either mention on its own. A plain "stated-once,
   ask-it-back" fact is NOT allowed - it must span distance.
2. Cross-segment information integration - the answer is what at least THREE
   facts stated in FAR-APART parts JOINTLY establish, never the facts
   themselves; drop any one of them and the answer no longer follows.
3. Multi-hop / multi-step inference - derive a fact no single statement
   supplies, by linking FOUR+ pieces spoken far apart, where each link only
   becomes usable once the previous one is established. Remove any one piece
   and the chain breaks; no piece and no halfway step gives it away alone.
4. Content-based speaker attribution - bind specific content to the RIGHT
   speaker across DISTANT parts, based on WHAT each one says. A single-segment
   "who said X" is NOT allowed.
5. Comparison / contrast of positions - relate how speakers view the same
   shared topic, using the positions each states in FAR-APART parts (never
   from a single exchange or adjacent turns).
6. Outcome / resolution interpretation - the answer is what the arc ultimately
   settles on once proposals, objections and revisions are tracked against
   each other. If one utterance announces the outcome outright, reframe it so
   the settled position only emerges from comparing proposed vs. survived.
7. Implicit inference - infer a meaning no speaker ever puts into words but
   that follows necessarily from several separate statements made in FAR-APART
   parts. The evidence is always spoken; only the conclusion is unsaid, so the
   answer can never rest on speculation, world knowledge, or an unvoiced detail.

==================================================================
E. HARD CONSTRAINTS (break any -> INVALID)
==================================================================
- CONTENT ONLY: never ask about tone, emotion, prosody, loudness, accent, or
  how anyone "sounded". Only what was SAID.
- NO TIMESTAMPS OR ORDERING CUES IN THE QUESTION OR OPTIONS: the "question"
  and "options" text must never mention seconds/minutes, nor position words
  that hint where evidence sits. Locating the relevant parts is the model's job.
  (This ban is ONLY for the question/options shown to the model. The audit
  fields "evidence" and "why_long_form" MUST carry approximate timestamps -
  see section J.) TEMPORAL/CAUSAL reasoning is still allowed: probe the RESULT
  of how events relate, without naming any position in the recording.
- TRUE DISTANT SEGMENTS ONLY: the segments the answer combines must come from
  NON-CONTIGUOUS parts of the recording - far apart on the timeline, separated
  by substantial intervening conversation (other turns/topics). Do NOT treat
  several consecutive sentences, one continuous stretch, or a single speaker
  turn as "multiple segments" - that is one segment, not distant parts, and
  makes the question INVALID. Each combined fact must sit in a genuinely
  different, distant region of the audio.
- NO SINGLE-SEGMENT ANSWERS: the answer must depend on multiple, distant parts.
- NO CLUE HANDOVER: the question may name its SUBJECT, but must not do the
  retrieval for the model. Never enumerate the specific facts, items or
  statements that must be connected, and never indicate where they occur.
- SPEAKER REFERENCES: never use names or speaker numbers. Refer to speakers by
  CONTENT. With {NUM_SPEAKERS} speakers, make attribution/comparison questions distinguish
  among all of them.
- SELF-CONTAINED: understandable to someone who only heard the audio.
- DIVERSITY: within each capability, the {N_PER_CAPABILITY} questions must use
  DIFFERENT parts/clues and have different answers - never rephrasings. Across
  capabilities, avoid reusing the same clue or answer.

==================================================================
F. OPTION & DISTRACTOR RULES
==================================================================
- Exactly TEN options, labelled (a)-(j). Exactly ONE defensibly correct; the
  other NINE are distractors.

CONSTRUCTION PROCEDURE (follow in order - do NOT invent options freely):
  1. The answer is a BINDING of two or more spoken facts (one fact bound to
     another, or a chain), never a single terminal value. If the answer is one
     spoken value with the rest of the option padded, the question is INVALID -
     redesign so the ANSWER ITSELF requires combining two or more facts.
  2. For each slot in the binding, list the REAL values actually spoken in the
     audio for that slot. ALL option parts come from THESE lists - never invent
     a value that was not spoken.
  3. Correct option = the true binding of the correct values.
  4. Each distractor = keep one slot correct and swap another slot for a
     DIFFERENT REAL value from the same list. Both halves stay individually TRUE
     and spoken; only the PAIRING is wrong. A model that extracted the facts but
     bound them incorrectly must land on a distractor.
  5. LEXICAL FLATTENING: every distinctive content word in the correct option
     must ALSO appear in at least three distractors. NO content word may be
     unique to the correct option - a word that sits only in the key lets the
     model keyword-match it to the audio and skip the reasoning entirely (the
     top failure mode). Rewrite any option set that leaves such a word.
- DERIVED CORRECT OPTION: the correct option states a value the model must
  DERIVE by combining distant segments, never a phrase quoted from the audio.
  Distractors are plausible but WRONG derivations - the result of extracting
  the right facts but combining them incorrectly, or deriving from a wrong
  subset. Extraction alone must land on a distractor; only correct derivation
  lands on the key.
- RELATION BINDING, NOT FACT RECOGNITION: the final challenge must be WHICH
  fact is bound to WHICH other fact - which cause with which effect, which
  remedy with which problem, which choice with which reason. Recognising the
  facts were mentioned must never be enough; only the correct relation is.
- CROSS-PRODUCT DISTRACTORS: form each distractor by keeping one side of the
  correct relation and swapping the other side for a DIFFERENT real fact from
  the recording. Both sides stay individually true - only the pairing is wrong.
- GROUNDED-ONLY: build every option from entities and facts EXPLICITLY spoken,
  so none can be dismissed as "not in the audio". This governs the building
  BLOCKS only; the correct option's RELATION/binding of those facts is still
  DERIVED, never a quoted phrase (no conflict with DERIVED CORRECT OPTION).
- NO COMMON-SENSE KILLS: never write an absurd or obviously irrelevant option.
  If an option can be rejected without listening, it does no work.

THE THREE WAYS A DISTRACTOR DIES (a dead distractor is a wasted option - it
silently shrinks a 10-way question to a 2-way or 3-way one). Every distractor
must survive ALL THREE tests, or REPLACE it:

  TEST 1 - ANSWER-FRAME CLOSURE (defeats the topic filter):
    Every fact you use as a filler must itself be a CANDIDATE ANSWER to the
    question asked. "Spoken somewhere in this recording" is NOT enough. If the
    question asks what triggers a condition, every filler must be something a
    speaker offered as a trigger for THAT condition - never a symptom of an
    unrelated condition, never a fact from a different topic in the same clip.
    A filler drawn from a different topic is eliminated by topic-matching
    alone, with the audio muted. Build ONE closed pool of same-frame candidate
    values per slot, and draw every option from that pool only.

  TEST 2 - SLOT INTERCHANGEABILITY (defeats world knowledge):
    Every filler must be TYPE-COMPATIBLE with EVERY slot it could occupy. Ask
    of each filler: "could this plausibly belong to any of the other slots, to
    someone who did not hear the recording?" If a value fits exactly one slot
    by its meaning alone, permuting it produces a distractor world knowledge
    rejects for free. Example of the FAILURE: pairing three remedies to three
    patient groups where each remedy is medically specific to one group - the
    binding is re-derivable without listening, so all nine distractors die.
    Fix it by choosing slots whose candidate values are mutually swappable
    (all three remedies could sensibly apply to any of the three groups), so
    ONLY the recording says which pairing actually occurred.

  TEST 3 - THE MIS-BINDING WITNESS (the positive test):
    For each distractor, name the specific listener error that produces it -
    "heard both facts but attached the later one to the earlier speaker",
    "integrated two of the three required segments", "followed the chain but
    stopped one hop short". Write that error in `distractor_rationale`. If you
    cannot name a plausible listener who lands there, the option is dead -
    replace it. Every one of the nine must be reachable by a DIFFERENT
    realistic error, so a partially-correct listener is spread across many
    wrong options rather than funnelled to the key.

- SELF-AUDIT BEFORE EMITTING: for each question, mute the audio in your head
  and answer using only the question text, the ten options, and world
  knowledge. If you can reach the key - or eliminate more than TWO options -
  the option set is too weak. Rebuild it under the three tests above. Target:
  a reader without the audio can eliminate NOTHING and must guess 1-in-10.
- MINIMAL PAIRS: at least 6 of 10 options differ from the correct one by
  exactly ONE fact or binding; at least 8 of 10 share the same sentence structure.
- EQUAL DETAIL BUDGET: every option asserts the SAME number of facts, in the
  SAME shape, with the SAME qualifiers and named specifics. Write the correct
  option FIRST, then build the others to that exact template, swapping only
  bindings. "FIRST" is a CONSTRUCTION step only; after all ten exist, PLACE the
  correct one at a RANDOM letter (see CORRECT-ANSWER POSITION), never at (a).
- SAME THEMATIC FAMILY (critical): all ten options sit in the SAME topic/domain
  family, differing ONLY in WHICH facts, speakers or details they assert. Never
  let each option name a different domain - that lets one recognised clue
  eliminate the other nine and collapses the chain to one hop.
- NO SURFACE CUES: the correct option must NOT be inferable from question
  wording, lexical overlap, or world knowledge. A quick/shallow read should
  point to a DISTRACTOR.
- DEEP CHAINS: where the capability allows, the answer should require chaining
  4+ scattered clues; prefer questions hinging on distinguishing NEAR-DUPLICATE
  mentions (similar things said at different points).
- No "all/none of the above"; no giveaway wording.
- CORRECT-ANSWER POSITION: the correct letter must be randomly distributed
  across (a)-(j) from question to question - do NOT cluster at any position.
- OPTION ORDERING: options sharing a similar descriptor must NOT be adjacent;
  shuffle so similar options are scattered, not grouped.

==================================================================
G. PER-CAPABILITY DISTRACTOR RECIPE
==================================================================
Every distractor is a WRONG BINDING of correct facts - but the AXIS you permute
is specific to the capability being tested, and must NOT be swapped for another
capability's axis. Permuting speaker identity everywhere, for instance, turns
all seven capabilities into speaker-attribution questions and destroys what each
one measures. Use the capability's own axis below, and apply the three
distractor-survival tests from section F to it.

- Long-context retention & recall : distractors = other near-duplicate details
  from distant points; force disambiguation by earlier context.
- Cross-segment information integration : distractors combine SOME but not all
  required facts (partial integration looks right).
- Multi-hop / multi-step inference : distractors = valid conclusions from a
  WRONG subset of clues (a defensible but incomplete chain).
- Content-based speaker attribution : distractors = SPEAKER-SWAPS (correct
  content attributed to the wrong participant); swap speakers in several options.
- Comparison / contrast of positions : distractors SWAP which speaker holds
  which position, or blend the two positions.
- Outcome / resolution interpretation : distractors = the REJECTED option, an
  intermediate step, or the outcome attributed to the wrong decider.
- Implicit inference : distractors = the surface-literal reading + an
  over-inference that goes one hop too far. All options name the SAME kind of
  theme, differing only in WHICH clues or speakers it covers.

==================================================================
H. SHORTCUT-DEFEAT VERIFICATION (every question must pass ALL)
==================================================================
1. EXTRACTION TEST: the answer is NOT spoken verbatim anywhere. If any single
   utterance states it, the question is INVALID - it must be DERIVED, not
   extracted (this is the primary gate; see section C, PILLAR 2).
2. SINGLE-SEGMENT TEST: no single spoken segment answers it. If one does, rewrite.
3. REMOVE-A-CLUE TEST: drop any one required clue -> the answer becomes
   genuinely undecidable between >=2 options. If one clue alone selects the
   answer, the question is INVALID.
4. ELIMINATION TEST: no option can be discarded without listening (all grounded,
   same thematic family, equal detail).
5. WORLD-KNOWLEDGE TEST: general/textbook knowledge alone cannot pick the answer.
6. KEYWORD-OVERLAP TEST: question words do not lexically point to the correct option.
7. AMBIGUITY CHECK: the correct answer is UNAMBIGUOUSLY supported and NO
   distractor is also defensibly correct. If two options could both be argued
   correct, discard and rewrite. Ambiguity is INVALID, not "hard".
8. GROUNDED-DISTRACTOR TEST: every value in every distractor was actually
   spoken in the audio. If any option contains an invented value (one never
   said), rewrite it - an unspoken option is discarded without reasoning.
9. LEXICAL-LEAK TEST: no content word appears in the correct option alone. If
   any distinctive word is unique to the key, the model keyword-matches it;
   spread that word across at least three distractors.
10. DERIVED-ANSWER TEST: the answer requires combining two or more spoken
   facts. If a single spoken fact selects the correct option, it is
   extraction, not derivation - redesign the option set.

==================================================================
I. LENGTH LIMITS
==================================================================
- QUESTION: max 15 words. No recap, no context, no described circumstance -
  every extra word is a clue that lets the model answer without the audio.
  Keep the logic HIDDEN. Cut context, never capability.
- OPTION: max 20 words each. A long option carries the answer inside it.

==================================================================
J. OUTPUT FORMAT (strict JSON array)
==================================================================
- "answer" MUST be ONLY the correct option letter in parentheses, e.g. "(c)" -
  no words, no option text.
- The answer MUST be DERIVED, never quoted or retrieved (see section C).
- "evidence" MUST list, SEPARATELY, each spoken piece the answer combines, and
  TAG EACH with its approximate timestamp in the recording, e.g.
  "[01:12] speaker who is studying says X; [07:48] the other says Y; [14:30] ...".
  The timestamps prove the pieces come from TRUE DISTANT segments (section E);
  if two tagged pieces are contiguous/adjacent, the question is INVALID.
- "why_long_form" MUST name which distant parts must connect AND include the
  timestamps of those connected segments, e.g. "needs [01:12]+[07:48]+[14:30];
  a clip hearing only one region cannot derive it".
- Timestamps appear ONLY in these two audit fields, NEVER in "question" or
  "options" (section E).

Produce EXACTLY {N_PER_CAPABILITY} questions for EACH capability below, in this
order, so the array has EXACTLY (7 x {N_PER_CAPABILITY}) objects. Do not skip,
merge, or reorder capabilities. The "capability" value is FIXED to its group.
(If {N_PER_CAPABILITY} > 1, repeat that capability's object that many times
before moving on.)

[
  {
    "capability": "Long-context retention & recall",
    "type": "mcq",
    "question": ".....",
    "options": "(a) ...\n(b) ...\n(c) ...\n(d) ...\n(e) ...\n(f) ...\n(g) ...\n(h) ...\n(i) ...\n(j) ...",
    "answer": "(X)",
    "distractor_rationale": "<one line: why the distractors are plausible but wrong>",
    "evidence": "[mm:ss] fact 1; [mm:ss] fact 2; [mm:ss] fact 3 - each from a TRUE DISTANT, non-contiguous part",
    "why_long_form": "needs [mm:ss]+[mm:ss]+[mm:ss]; a clip hearing only one region cannot derive it"
  },
  {
    "capability": "Cross-segment information integration",
    "type": "mcq",
    "question": ".....",
    "options": "(a) ...\n(b) ...\n(c) ...\n(d) ...\n(e) ...\n(f) ...\n(g) ...\n(h) ...\n(i) ...\n(j) ...",
    "answer": "(X)",
    "distractor_rationale": "<one line: why the distractors are plausible but wrong>",
    "evidence": "[mm:ss] fact 1; [mm:ss] fact 2; [mm:ss] fact 3 - each from a TRUE DISTANT, non-contiguous part",
    "why_long_form": "needs [mm:ss]+[mm:ss]+[mm:ss]; a clip hearing only one region cannot derive it"
  },
  {
    "capability": "Multi-hop / multi-step inference",
    "type": "mcq",
    "question": ".....",
    "options": "(a) ...\n(b) ...\n(c) ...\n(d) ...\n(e) ...\n(f) ...\n(g) ...\n(h) ...\n(i) ...\n(j) ...",
    "answer": "(X)",
    "distractor_rationale": "<one line: why the distractors are plausible but wrong>",
    "evidence": "[mm:ss] fact 1; [mm:ss] fact 2; [mm:ss] fact 3 - each from a TRUE DISTANT, non-contiguous part",
    "why_long_form": "needs [mm:ss]+[mm:ss]+[mm:ss]; a clip hearing only one region cannot derive it"
  },
  {
    "capability": "Content-based speaker attribution",
    "type": "mcq",
    "question": ".....",
    "options": "(a) ...\n(b) ...\n(c) ...\n(d) ...\n(e) ...\n(f) ...\n(g) ...\n(h) ...\n(i) ...\n(j) ...",
    "answer": "(X)",
    "distractor_rationale": "<one line: why the distractors are plausible but wrong>",
    "evidence": "[mm:ss] fact 1; [mm:ss] fact 2; [mm:ss] fact 3 - each from a TRUE DISTANT, non-contiguous part",
    "why_long_form": "needs [mm:ss]+[mm:ss]+[mm:ss]; a clip hearing only one region cannot derive it"
  },
  {
    "capability": "Comparison / contrast of positions",
    "type": "mcq",
    "question": ".....",
    "options": "(a) ...\n(b) ...\n(c) ...\n(d) ...\n(e) ...\n(f) ...\n(g) ...\n(h) ...\n(i) ...\n(j) ...",
    "answer": "(X)",
    "distractor_rationale": "<one line: why the distractors are plausible but wrong>",
    "evidence": "[mm:ss] fact 1; [mm:ss] fact 2; [mm:ss] fact 3 - each from a TRUE DISTANT, non-contiguous part",
    "why_long_form": "needs [mm:ss]+[mm:ss]+[mm:ss]; a clip hearing only one region cannot derive it"
  },
  {
    "capability": "Outcome / resolution interpretation",
    "type": "mcq",
    "question": ".....",
    "options": "(a) ...\n(b) ...\n(c) ...\n(d) ...\n(e) ...\n(f) ...\n(g) ...\n(h) ...\n(i) ...\n(j) ...",
    "answer": "(X)",
    "distractor_rationale": "<one line: why the distractors are plausible but wrong>",
    "evidence": "[mm:ss] fact 1; [mm:ss] fact 2; [mm:ss] fact 3 - each from a TRUE DISTANT, non-contiguous part",
    "why_long_form": "needs [mm:ss]+[mm:ss]+[mm:ss]; a clip hearing only one region cannot derive it"
  },
  {
    "capability": "Implicit inference",
    "type": "mcq",
    "question": ".....",
    "options": "(a) ...\n(b) ...\n(c) ...\n(d) ...\n(e) ...\n(f) ...\n(g) ...\n(h) ...\n(i) ...\n(j) ...",
    "answer": "(X)",
    "distractor_rationale": "<one line: why the distractors are plausible but wrong>",
    "evidence": "[mm:ss] fact 1; [mm:ss] fact 2; [mm:ss] fact 3 - each from a TRUE DISTANT, non-contiguous part",
    "why_long_form": "needs [mm:ss]+[mm:ss]+[mm:ss]; a clip hearing only one region cannot derive it"
  }
]

Return ONLY the JSON array. No commentary.
\end{tcblisting}

\begin{tcblisting}{listing only, breakable, enhanced,
  colback=black!3, colframe=black!55, boxrule=0.4pt, arc=1mm,
  left=2mm, right=2mm, top=1mm, bottom=1mm,
  title={Long-Form Audio Reasoning: Open-Ended QA Generation Prompt}, fonttitle=\bfseries\small,
  listing options={basicstyle=\scriptsize\ttfamily, breaklines=true,
    columns=fullflexible, keepspaces=true}}
==================================================================
A. ROLE & GOAL
==================================================================
You are an expert examiner building an ADVERSARIAL benchmark for LONG-FORM
AUDIO REASONING in audio LLMs. You are given the AUDIO of ONE complete
multi-speaker conversation. Listen to the ENTIRE recording, then write
OPEN-ENDED question-answer pairs (free-text answer).

Your GOAL is to create questions that FAIL any model taking a SHORTCUT -
single-segment lookup, keyword matching, or world knowledge - while staying
UNAMBIGUOUSLY correct for a listener who truly integrated the whole
conversation. Above all, each question must force the model to DERIVE the
answer by combining multiple far-distant spoken segments, never to extract or
recall a value spoken aloud. Questions that defeat shortcut-reliant models are
what make the benchmark strong.

Assume the model under test is a STRONG audio LLM: excellent at understanding
any single segment, with broad world knowledge and sharp pattern-matching.
Your questions must be hard enough that even this strong model FAILS unless it
genuinely tracked and integrated the WHOLE conversation and DERIVED the answer
across distant segments. The one thing it may lack - and what you are probing -
is long-form derivation. Assume it will attempt every shortcut (single-segment
lookup, keyword match, world knowledge); design each question so those
shortcuts land on a WRONG answer and only whole-conversation derivation reaches
the correct one. Every question you write is correct and defensible from the
audio. Difficulty comes ONLY from how far apart and how well hidden the evidence
is, and from the derivation the answer demands - NEVER from ambiguity or a
wrong ground truth.

OPEN-ENDED NOTE: with no options to eliminate, the answer must be a value the
model GENERATES. This makes the single-ground-truth requirement stricter, not
looser: the derived value must be specific enough that a grader can mark a
free-text response right or wrong without judgement calls.

==================================================================
B. INPUT
==================================================================
- AUDIO: one full multi-speaker conversation (near-field, spontaneous, unscripted).
- NUMBER_OF_SPEAKERS: {NUM_SPEAKERS}
- NUMBER_OF_QUESTIONS_PER_CAPABILITY: {N_PER_CAPABILITY}
- QUESTION_LANGUAGE: {QUESTION_LANGUAGE}
You have no transcript and no timestamps - rely only on what you HEAR. Base
every question and answer strictly on the SPOKEN CONTENT of this recording.

==================================================================
C. HOW TO FORM A QUESTION (two pillars: cross-segment integration + DERIVATION)
==================================================================
Every question must FORCE the model to DERIVE (generate) the answer, not
extract or retrieve it. Both pillars are required on EVERY question:

PILLAR 1 - CROSS-SEGMENT INTEGRATION (where the evidence lives):
  The evidence is scattered across MULTIPLE, potentially far-distant segments.
  No single segment carries the answer. Remove any one required segment and
  the answer becomes underdetermined. The model must LOCATE the relevant
  segments itself - do not point to where they are; finding them is the test.

PILLAR 2 - DERIVATION, NOT EXTRACTION (what the answer is):
  The answer is a value the model must WORK OUT by combining those segments -
  a relation, a resolution, an inference, a computed conclusion - that NO
  speaker ever states aloud. If the answer equals something spoken in the
  audio, it is a simple information-extraction question and is INVALID.
  Derivation is the point: recognising or recalling facts is never enough; the
  model must GENERATE the answer from the combination of scattered evidence.

RELATIONSHIP TYPES: whichever capability a question tests, its derivation
connects temporally distant, distributed evidence through one or more of these
relationships - TEMPORAL, CAUSAL, CROSS-SPEAKER, REFERENTIAL, or COMPARATIVE.
The answer is derived by relating evidence across far-apart segments, never
read from any single one.

Why shortcuts fail: any model that stops at extraction, or connects the WRONG
segments, answers incorrectly - no matter how strong its single-segment
perception or world knowledge. Only genuine derivation over the whole
conversation yields the correct, defensible answer.

==================================================================
D. CAPABILITIES TO TEST (tag every question with exactly one capability)
==================================================================
1. Long-context retention & recall - the answer is the LINK between an early
   mention and a distant later one (what it revises, contradicts or
   disambiguates), never either mention on its own. A plain "stated-once,
   ask-it-back" fact is NOT allowed - it must span distance.
2. Cross-segment information integration - the answer is what at least THREE
   facts stated in FAR-APART parts JOINTLY establish, never the facts
   themselves; drop any one of them and the answer no longer follows.
3. Multi-hop / multi-step inference - derive a fact no single statement
   supplies, by linking FOUR+ pieces spoken far apart, where each link only
   becomes usable once the previous one is established. Remove any one piece
   and the chain breaks; no piece and no halfway step gives it away alone.
4. Content-based speaker attribution - bind specific content to the RIGHT
   speaker across DISTANT parts, based on WHAT each one says. A single-segment
   "who said X" is NOT allowed.
5. Comparison / contrast of positions - relate how speakers view the same
   shared topic, using the positions each states in FAR-APART parts (never
   from a single exchange or adjacent turns).
6. Outcome / resolution interpretation - the answer is what the arc ultimately
   settles on once proposals, objections and revisions are tracked against
   each other. If one utterance announces the outcome outright, reframe it so
   the settled position only emerges from comparing proposed vs. survived.
7. Implicit inference - infer a meaning no speaker ever puts into words but
   that follows necessarily from several separate statements made in FAR-APART
   parts. The evidence is always spoken; only the conclusion is unsaid, so the
   answer can never rest on speculation, world knowledge, or an unvoiced detail.

==================================================================
E. HARD CONSTRAINTS (break any -> INVALID)
==================================================================
- CONTENT ONLY: never ask about tone, emotion, prosody, loudness, accent, or
  how anyone "sounded". Only what was SAID.
- NO TIMESTAMPS OR ORDERING CUES IN THE QUESTION: the "question" text must
  never mention seconds/minutes, nor position words that hint where evidence
  sits. Locating the relevant parts is the model's job.
  (This ban is ONLY for the question shown to the model. The audit fields
  "evidence" and "why_long_form" MUST carry approximate timestamps - see
  section J.) TEMPORAL/CAUSAL reasoning is still allowed: probe the RESULT of
  how events relate, without naming any position in the recording.
- TRUE DISTANT SEGMENTS ONLY: the segments the answer combines must come from
  NON-CONTIGUOUS parts of the recording - far apart on the timeline, separated
  by substantial intervening conversation (other turns/topics). Do NOT treat
  several consecutive sentences, one continuous stretch, or a single speaker
  turn as "multiple segments" - that is one segment, not distant parts, and
  makes the question INVALID. Each combined fact must sit in a genuinely
  different, distant region of the audio.
- NO SINGLE-SEGMENT ANSWERS: the answer must depend on multiple, distant parts.
- NO CLUE HANDOVER: the question may name its SUBJECT, but must not do the
  retrieval for the model. Never enumerate the specific facts, items or
  statements that must be connected, and never indicate where they occur.
- SPEAKER REFERENCES: never use names or speaker numbers. Refer to speakers by
  CONTENT ("the speaker who is studying", "one participant / the other"). With
  {NUM_SPEAKERS} speakers, make attribution/comparison questions distinguish
  among all of them.
- SELF-CONTAINED: understandable to someone who only heard the audio.
- DIVERSITY: within each capability, the {N_PER_CAPABILITY} questions must use
  DIFFERENT parts/clues and have different answers - never rephrasings. Across
  capabilities, avoid reusing the same clue or answer.

==================================================================
F. ANSWER RULES
==================================================================
The ANSWER itself carries the whole difficulty. Build
it under these rules:

CONSTRUCTION PROCEDURE (follow in order):
  1. The answer is a value DERIVED by combining two or more facts spoken in
     far-apart parts. It may be a single specific value (a duration, a count,
     a resolved reference, a settled position) - what matters is that NO
     speaker states it aloud. If any single utterance supplies it, the
     question is INVALID; redesign so reaching it REQUIRES combining separate
     facts.
  2. Identify each fact the derivation depends on. Then check: if a listener
     caught only SOME of them, or connected them wrongly, what answer would
     they give? Name at least THREE such plausible wrong answers.
  3. If mis-integration produces nothing plausible - if a partial listener
     simply has no answer - the question is too easy or too obscure; redesign
     it. The wrong answers are a DESIGN CHECK only; never write them into the
     output.
  4. GRADABILITY: the correct answer must be a SPECIFIC, checkable value or
     relation - short, committed, and phrased so that a free-text response
     either matches it or does not. Never an open-ended description, a
     narrative, or a summary that a grader must judge subjectively.
- DERIVED ANSWER: the answer states a value the model must DERIVE by combining
  distant segments, never a phrase quoted from the audio. Extraction alone must
  produce a WRONG answer; only correct derivation reaches the right one.
- RELATION, NOT FACT RECOGNITION: the final challenge must be the RELATION
  between facts, not the facts themselves - which cause goes with which
  effect, which choice with which reason, what a chain of statements adds up
  to. Recognising the facts were mentioned must never be enough; only
  correctly relating them is.
- GROUNDED-ONLY: build the answer from entities and facts EXPLICITLY spoken.
  This governs the building BLOCKS only; the answer's RELATION/binding of those
  facts is still DERIVED, never a quoted phrase.
- SINGLE GROUND TRUTH, TEMPTING ALTERNATIVES: each question must have exactly
  ONE answer that is correct given the audio. It is GOOD - even preferred - for
  several plausible-sounding answers to exist, as long as only one is actually
  supported and the rest are subtly wrong (true of the other speaker, half-true,
  or one hop short). Do NOT write genuinely ambiguous questions where two or
  more DIFFERENT answers would both be correct - those cannot be graded.
- NO ANSWER-AS-EVIDENCE-LIST: the answer is the derived value itself, never a
  recital of the facts it rests on. Those go in "evidence", listed separately.
- SELF-AUDIT BEFORE EMITTING: for each question, mute the audio in your head
  and try to answer using only the question text and world knowledge. If you
  can reach the correct answer, or narrow it to a small set, the question is too
  weak - rebuild it. Target: a reader without the audio cannot even guess.

==================================================================
G. PER-CAPABILITY TRAP (which mis-binding the question must invite)
==================================================================
Every wrong answer is a WRONG BINDING of correct facts - but the AXIS the model
is invited to mis-permute is specific to the capability being tested, and must
NOT be swapped for another capability's axis. Inviting a speaker-identity error
everywhere, for instance, turns all seven capabilities into speaker-attribution
questions and destroys what each one measures.

- Long-context retention & recall : the trap is another near-duplicate detail
  from a distant point; the answer hinges on disambiguating by earlier context.
- Cross-segment information integration : the trap combines SOME but not all
  required facts (partial integration looks right).
- Multi-hop / multi-step inference : the trap is a valid conclusion from a
  WRONG subset of clues (a defensible but incomplete chain).
- Content-based speaker attribution : the trap is a SPEAKER-SWAP (correct
  content attributed to the wrong participant).
- Comparison / contrast of positions : the trap SWAPS which speaker holds which
  position, or blends the two positions.
- Outcome / resolution interpretation : the trap is the REJECTED proposal, an
  intermediate step, or the outcome attributed to the wrong decider.
- Implicit inference : the trap is the surface-literal reading or an
  over-inference that goes one hop too far.

==================================================================
H. SHORTCUT-DEFEAT VERIFICATION (every question must pass ALL)
==================================================================
1. EXTRACTION TEST: the answer is NOT spoken verbatim anywhere. If any single
   utterance states it, the question is INVALID - it must be DERIVED, not
   extracted (this is the primary gate; see section C, PILLAR 2).
2. SINGLE-SEGMENT TEST: no single spoken segment answers it. If one does, rewrite.
3. REMOVE-A-CLUE TEST: drop any one required clue -> the answer becomes
   genuinely undecidable. If one clue alone selects the answer, the question is
   INVALID.
4. DISTANCE TEST: the required pieces sit in non-contiguous, far-apart regions.
   If two are adjacent or in one turn, the question is INVALID (section E).
5. WORLD-KNOWLEDGE TEST: general/textbook knowledge alone cannot produce the answer.
6. KEYWORD-OVERLAP TEST: question words do not lexically point to the answer.
7. AMBIGUITY CHECK: the answer is UNAMBIGUOUSLY supported and no DIFFERENT
   answer is also defensibly correct. If two answers could both be argued
   correct, discard and rewrite. Ambiguity is INVALID, not "hard".
8. GRADABILITY TEST: the answer is specific and committed enough that a grader
   comparing a free-text response against it can decide right/wrong without a
   judgement call. If grading would be subjective, rewrite.
9. DERIVED-ANSWER TEST: the answer requires combining two or more spoken facts.
   If a single spoken fact yields it, it is extraction, not derivation -
   redesign the question.

==================================================================
I. LENGTH LIMITS
==================================================================
- QUESTION: max 15 words. No recap, no context, no described circumstance -
  every extra word is a clue that lets the model answer without the audio.
  Keep the logic HIDDEN. Cut context, never capability.
  Count the words, then ask of each one left: would it help answer WITHOUT the
  audio? If yes, delete it.
- ANSWER: keep it concise and committed - the derived value itself. The 15-word
  cap is on the QUESTION only; the answer may run longer where the derived
  relation genuinely needs it, but never becomes a narrative or a recital of
  the evidence.

==================================================================
J. OUTPUT FORMAT (strict JSON array)
==================================================================
- "answer" MUST be the derived ground-truth value itself - free text, no option
  letters, no quoted utterance.
- The answer MUST be DERIVED, never quoted or retrieved (see section C).
- "evidence" MUST list, SEPARATELY, each spoken piece the answer combines, and
  TAG EACH with its approximate timestamp in the recording, e.g.
  "[01:12] speaker who is studying says X; [07:48] the other says Y; [14:30] ...".
  The timestamps prove the pieces come from TRUE DISTANT segments (section E);
  if two tagged pieces are contiguous/adjacent, the question is INVALID.
- "why_long_form" MUST name which distant parts must connect AND include the
  timestamps of those connected segments, e.g. "needs [01:12]+[07:48]+[14:30];
  a clip hearing only one region cannot derive it".
- Timestamps appear ONLY in these two audit fields, NEVER in "question"
  (section E).

Produce EXACTLY {N_PER_CAPABILITY} questions for EACH capability below, in this
order, so the array has EXACTLY (7 x {N_PER_CAPABILITY}) objects. Do not skip,
merge, or reorder capabilities. The "capability" value is FIXED to its group.
(If {N_PER_CAPABILITY} > 1, repeat that capability's object that many times
before moving on.)

[
  {
    "capability": "Long-context retention & recall",
    "type": "open",
    "question": ".....",
    "answer": "<the concise DERIVED ground-truth answer>",
    "evidence": "[mm:ss] fact 1; [mm:ss] fact 2; [mm:ss] fact 3 - each from a TRUE DISTANT, non-contiguous part",
    "why_long_form": "needs [mm:ss]+[mm:ss]+[mm:ss]; a clip hearing only one region cannot derive it"
  },
  {
    "capability": "Cross-segment information integration",
    "type": "open",
    "question": ".....",
    "answer": "<the concise DERIVED ground-truth answer>",
    "evidence": "[mm:ss] fact 1; [mm:ss] fact 2; [mm:ss] fact 3 - each from a TRUE DISTANT, non-contiguous part",
    "why_long_form": "needs [mm:ss]+[mm:ss]+[mm:ss]; a clip hearing only one region cannot derive it"
  },
  {
    "capability": "Multi-hop / multi-step inference",
    "type": "open",
    "question": ".....",
    "answer": "<the concise DERIVED ground-truth answer>",
    "evidence": "[mm:ss] fact 1; [mm:ss] fact 2; [mm:ss] fact 3 - each from a TRUE DISTANT, non-contiguous part",
    "why_long_form": "needs [mm:ss]+[mm:ss]+[mm:ss]; a clip hearing only one region cannot derive it"
  },
  {
    "capability": "Content-based speaker attribution",
    "type": "open",
    "question": ".....",
    "answer": "<the concise DERIVED ground-truth answer>",
    "evidence": "[mm:ss] fact 1; [mm:ss] fact 2; [mm:ss] fact 3 - each from a TRUE DISTANT, non-contiguous part",
    "why_long_form": "needs [mm:ss]+[mm:ss]+[mm:ss]; a clip hearing only one region cannot derive it"
  },
  {
    "capability": "Comparison / contrast of positions",
    "type": "open",
    "question": ".....",
    "answer": "<the concise DERIVED ground-truth answer>",
    "evidence": "[mm:ss] fact 1; [mm:ss] fact 2; [mm:ss] fact 3 - each from a TRUE DISTANT, non-contiguous part",
    "why_long_form": "needs [mm:ss]+[mm:ss]+[mm:ss]; a clip hearing only one region cannot derive it"
  },
  {
    "capability": "Outcome / resolution interpretation",
    "type": "open",
    "question": ".....",
    "answer": "<the concise DERIVED ground-truth answer>",
    "evidence": "[mm:ss] fact 1; [mm:ss] fact 2; [mm:ss] fact 3 - each from a TRUE DISTANT, non-contiguous part",
    "why_long_form": "needs [mm:ss]+[mm:ss]+[mm:ss]; a clip hearing only one region cannot derive it"
  },
  {
    "capability": "Implicit inference",
    "type": "open",
    "question": ".....",
    "answer": "<the concise DERIVED ground-truth answer>",
    "evidence": "[mm:ss] fact 1; [mm:ss] fact 2; [mm:ss] fact 3 - each from a TRUE DISTANT, non-contiguous part",
    "why_long_form": "needs [mm:ss]+[mm:ss]+[mm:ss]; a clip hearing only one region cannot derive it"
  }
]

Return ONLY the JSON array. No commentary.
\end{tcblisting}

\begin{tcblisting}{listing only, breakable, enhanced,
  colback=black!3, colframe=black!55, boxrule=0.4pt, arc=1mm,
  left=2mm, right=2mm, top=1mm, bottom=1mm,
  title={Prosodic Ambiguity Resolution (V1): QA Generation Prompt}, fonttitle=\bfseries\small,
  listing options={basicstyle=\scriptsize\ttfamily, breaklines=true,
    columns=fullflexible, keepspaces=true}}
You are a meticulous benchmark author. You write ONE question that tests whether a listener can
resolve a STRUCTURALLY AMBIGUOUS spoken Indonesian sentence using ONLY the audio (prosody: stress,
pauses, intonation). In writing the sentence has two grammatically valid readings; the recording
commits to exactly one of them. THE RECORDING ITSELF IS ATTACHED as audio input - LISTEN to it
before writing anything. You are also given the sentence and both readings in English, plus the
GOLD reading THIS recording expresses. The correct answer is FIXED by the GOLD - never pick based on
which reading seems more plausible in general.

Follow the given qa_format:

- "mcq": Write a focused question about the SINGLE point of ambiguity, then exactly {{num_options}}
  answer options (letters A-J). This is a HARD benchmark: the option set must make every shortcut
  fail. Requirements:
    * Exactly ONE option matches the GOLD reading (the correct answer).
    * ONE option expresses the OTHER valid reading (the hardest distractor).
    * ONE option is the safe-play escape: "The recording doesn't say" (or a natural equivalent).
      It must be WRONG - the audio does disambiguate.
    * Build the REMAINING options as the strongest wrong answers you can. HARD-DISTRACTOR rules:
        - prefer WRONG BINDINGS of entities/actions/locations that ARE in the sentence (swap
          which noun the property applies to, attach the phrase to the wrong action, invert
          who does what) - these force real parsing of the audio;
        - "both X and Y" / "neither X nor Y" mis-scopings;
        - subtle corruptions of the gold reading (right entity, wrong action; right action,
          wrong entity);
        - at most ONE option may introduce content not present in the sentence at all (besides
          the safe-play option) - foreign-content options are too easy to eliminate;
        - a distractor must NEVER amount to a third grammatically valid reading of the sentence
          - there must remain exactly one defensible answer.
    * DEDUPLICATION: every option must assert a DISTINCT state of affairs. No two options may
      differ only by a synonym, degree word (well/thoroughly), or rewording of the same claim.
      In particular, no two options may express the SAME underlying reading of the sentence in
      different words - each of the two valid readings appears exactly ONCE in the option set.
    * Every option must be a direct, well-formed answer to the question - no meta options like
      "all of the above", no jokes.
    * Keep options concise, mutually exclusive, and similar in length/style so wording never leaks
      the answer.
    * Vary which letter is correct across items; do not default to a fixed position.
  Distractor guidance by ambiguity type:
    * Relative-clause / prepositional-phrase / verb attachment (Type04/05/06): "both" or "neither"
      are acceptable distractors.
    * Coordination / modifier scope (Type10): one of the two REAL readings already means "both ...",
      so NEVER use "both" as a throwaway distractor here - use "only the other item", "neither",
      or an unstated-entity distractor instead.

- "open_ended": Write a focused question asking the listener to state the intended meaning. The
  question must be phrased so that EITHER reading's distinguishing phrase would be a natural and
  complete answer to it - it must not fit the gold better than the other reading. Set
  reference_answer to the GOLD distinguishing phrase COPIED VERBATIM - do not paraphrase, expand,
  or reword it.

Question phrasing: use natural, varied wording - there is NO required template or fixed opening.
The only requirements are that the question is clearly about what the speaker MEANS in the audio
(not about general world knowledge) and that it targets the point of ambiguity. Vary the phrasing
style across items.

CRITICAL - no transcript leakage: NEVER quote the sentence, reproduce it in translation, or closely
paraphrase its full content inside the question (or inside any option). The listener must get the
words from the AUDIO, not from you. Refer only to the minimal entities/events needed to pose the
question (e.g. "what did the speaker pick up at the beach?" is fine; "in the sentence '...'" is
forbidden). Additionally: do NOT quote ANY words or phrases from the sentence (no quotation marks
around its wording at all), and do NOT use meta-linguistic pointers such as "the phrase X", "the
prepositional phrase", or "the modifier" - pose the question purely in terms of meaning. The
question must never assert facts the listener is supposed to extract by listening.

CRITICAL - frame neutrality (no answer bias): the question must be answerable under BOTH readings,
with DIFFERENT answers. It must NOT be built from the gold's event frame, and the gold must not be
guessable from the question text alone. Forbidden pattern: asking "where/when/how does <the gold
reading's action> happen?" when, under the other reading, that question would have no answer or the
answer "unspecified" (e.g. for a sentence ambiguous between "he resells it in the market" and "he
bought it in the market", the question "where does the reselling take place?" is FORBIDDEN - it
presupposes the gold; ask instead "which action does the speaker say happened in the market?").
SELF-CHECK before finalizing: silently answer your question under reading 1 and under reading 2.
If either reading yields no answer, "unspecified", or the SAME answer as the other reading,
REWRITE the question. This rule applies to BOTH mcq and open_ended.

Coordination-scope phrasing (Type10): when the two readings differ on whether a property applies
to ONE item or BOTH items, the question must ask "WHICH item(s) ..." and let the listener choose
the subset. NEVER jointly predicate both items in the question (e.g. "what is said about the
condition of A and B?" is FORBIDDEN - it fits the both-reading better; ask "which of the stolen
items are described as new?" style instead).

Format example (ILLUSTRATION ONLY - do not reuse its wording, topic, sentence pattern, or option
style; your item must be built from the sentence you are given):
  Question: "What did the speaker personally pick up at the beach?"
  Options:  A) The sand   B) The shell   C) Both the sand and the shell   D) The recording
  doesn't say   ... (continues through J with strong wrong answers)

REQUIRED FIELD - GT_audio_reasoning (both formats): after listening to the attached recording,
write a 5-6 sentence ground-truth audio reasoning trace grounded in THIS specific recording:
(1) name the structural ambiguity and the two candidate readings; (2) describe the concrete
acoustic cues you actually perceive - where the prosodic boundary or pause falls, which word
carries stress/emphasis or lengthening, how the phrases are grouped, the pacing; (3) explain how
those cues select the GOLD reading over the other one. Describe ONLY cues you can genuinely
perceive in the audio - NEVER invent or assume cues; if the audio does not clearly support the
gold reading, state that explicitly inside the trace. Do not reference option letters - the trace
must stand alone as the reference reasoning for this clip.

Write the question and options in {{question_language}}, referring to entities by their English
names. Output STRICT JSON only - no markdown, no code fences, no commentary.

The recording is attached as audio input - listen to it first.

qa_format: {{qa_format}}
ambiguity_type: {{ambiguity_type}} - {{ambiguity_type_label}}
ambiguity_description: {{ambiguity_description}}

Spoken sentence (English): {{transcript_en}}
Reading 1 (English): {{interpretation_1_en}}
Reading 2 (English): {{interpretation_2_en}}
GOLD reading expressed by THIS recording (CORRECT answer): {{intended_interpretation_en}}
GOLD distinguishing phrase: {{gold_answer_en}}

Output JSON schema (use the one matching qa_format):
  mcq        -> {"question": str, "options": {"A": str, "B": str, ..., "J": str},   // exactly {{num_options}} options A-J
                 "answer": "A-J letter", "answer_text": str, "distractor_letters": [str, ...],
                 "GT_audio_reasoning": str,   // 5-6 sentence audio reasoning trace from THIS recording
                 "rationale": str}
  open_ended -> {"question": str, "reference_answer": str,
                 "GT_audio_reasoning": str,   // 5-6 sentence audio reasoning trace from THIS recording
                 "rationale": str}

{{type10_reminder}}Return only the JSON object.
\end{tcblisting}

\begin{tcblisting}{listing only, breakable, enhanced,
  colback=black!3, colframe=black!55, boxrule=0.4pt, arc=1mm,
  left=2mm, right=2mm, top=1mm, bottom=1mm,
  title={Prosodic Ambiguity Resolution (V2): QA Generation Prompt}, fonttitle=\bfseries\small,
  listing options={basicstyle=\scriptsize\ttfamily, breaklines=true,
    columns=fullflexible, keepspaces=true}}
You are a meticulous benchmark author. A single audio clip contains the SAME Indonesian sentence
spoken TWICE by the same speaker. The words are identical both times; ONLY the prosody differs, so
each utterance expresses a DIFFERENT one of the sentence's two valid readings. THE CLIP ITSELF IS
ATTACHED as audio input - LISTEN to both utterances before writing anything. Your question tests
whether a listener can tell, from the audio alone, which reading EACH utterance conveys.

You are given both readings in English and, for the FIRST and the SECOND utterance (in the order they
are heard in the clip), the GOLD reading each one expresses. The correct answer is FIXED by these two
golds - never infer it.

IMPORTANT - what the question may reveal: the question MUST begin with this exact sentence:
"The given audio has two different utterances."
Give NO other structural information: do NOT reveal that the two utterances are the same sentence,
and do not describe pauses, speakers, or recording details. After that fixed opening sentence, refer
to the utterances only as "the first utterance" and "the second utterance".

Follow the given qa_format:

- "mcq": After the fixed opening sentence, ask what the speaker means in the first utterance and in
  the second utterance. The question itself must make clear that each option lists the two meanings
  IN ORDER (first utterance, then second utterance). Provide exactly {{num_options}} options
  (letters A-J), each a COMPACT ordered pair of meanings separated by a comma:
      "<meaning in first utterance>, <meaning in second utterance>"
  Do NOT prefix options with labels like "First utterance:" - keep them short.
  This is a HARD benchmark. Build the option set as follows:
    * The 4 pairings of the two REAL readings: correct mapping, swapped mapping, and the two
      same-both-times mappings. EXACTLY ONE of these - (first -> FIRST GOLD, second -> SECOND GOLD) -
      is the correct answer; set "answer" to its letter.
    * ONE safe-play escape option such as "The audio doesn't make either meaning clear".
      It must be WRONG - the prosody does disambiguate.
    * Build the REMAINING options from strong wrong meanings. HARD-DISTRACTOR rules:
        - prefer pairings built from WRONG BINDINGS of entities/actions that ARE in the sentence
          (wrong attachment, wrong scope, inverted roles) - these force real parsing of the audio;
        - subtle corruptions of a real reading (right entity, wrong action; right action, wrong
          entity) are good;
        - at most ONE option may involve content not present in the sentence at all (besides the
          escape option) - foreign-content options are too easy to eliminate;
        - an invented meaning must NEVER amount to a third grammatically valid reading of the
          sentence - there must remain exactly one defensible answer.
    * DEDUPLICATION: every option must assert a DISTINCT pair of meanings. No two options may
      differ only by a synonym, degree word, or rewording of the same claim.
    * Every option must follow the same parallel two-part structure (except the escape option),
      be mutually exclusive with the others, and be similar in length so wording never leaks the
      answer.
    * Vary which letter is correct across items; do not default to a fixed position.

- "open_ended": After the fixed opening sentence, ask the listener to describe what the speaker
  means in the first utterance and in the second utterance. The question must be phrased so that
  EITHER reading's key phrase would be a natural and complete answer for either utterance - it
  must not fit one reading, or one ordering, better than the other. Set reference_answer_first and
  reference_answer_second to the respective GOLD key phrases COPIED VERBATIM - do not paraphrase,
  expand, or reword them.

Question phrasing: apart from the mandatory opening sentence, use natural, varied wording - no
fixed template. Vary the phrasing style across items.

CRITICAL - no transcript leakage: NEVER quote the sentence, reproduce it in translation, or closely
paraphrase its full content inside the question. Quoting it would also reveal that the two
utterances share the same words - structural information that must stay hidden. Refer only to the
minimal entities/events needed to pose the question. Additionally: do NOT quote ANY words or
phrases from the sentence (no quotation marks around its wording at all), and do NOT use
meta-linguistic pointers such as "the phrase X", "the prepositional phrase", or "the modifier" -
pose the question purely in terms of meaning. The question must never assert facts the listener
is supposed to extract by listening.

CRITICAL - frame neutrality (no answer bias): the question stem must be strictly neutral between
the two readings AND between the two possible orderings. The stem must target the actual point of
CONTRAST between the two readings (e.g. WHICH items / WHO / WHICH action), never a dimension on
which both readings agree (asking "where...?" when both readings name the same place is wrong). It must NOT be built from either
reading's event frame, must not hint which meaning comes first, and the correct pairing must not
be guessable from the question text alone. SELF-CHECK before finalizing: your question must read
identically sensibly if the two utterances' meanings were swapped; if swapping would make the
question fit worse, REWRITE it. This rule applies to BOTH mcq and open_ended.

Format example (ILLUSTRATION ONLY - do not reuse its wording, topic, or sentence pattern; your item
must be built from the sentence you are given):
  Question: "The given audio has two different utterances. What did the speaker personally pick up
  at the beach, in the first utterance and in the second utterance respectively?"
  Options:  A) The sand, The shell   B) The shell, The sand   C) The sand, The sand
            D) The shell, The shell   ... (continues through J with strong wrong pairings and one
            escape option)

REQUIRED FIELD - GT_audio_reasoning (both formats): after listening to the attached clip, write a
5-6 sentence ground-truth audio reasoning trace grounded in THIS specific clip, covering BOTH
utterances in the order they are heard: (1) note the two utterances share the same words and name
the two candidate readings; (2) for the FIRST utterance, describe the concrete acoustic cues you
actually perceive (prosodic boundary/pause placement, stressed or lengthened words, phrase
grouping, pacing) and how they select its GOLD reading; (3) do the same for the SECOND utterance,
contrasting what changed acoustically between the two. Describe ONLY cues you can genuinely
perceive - NEVER invent or assume cues; if the audio does not clearly support a gold reading,
state that explicitly inside the trace. Do not reference option letters - the trace must stand
alone as the reference reasoning for this clip.

Write in {{question_language}}, referring to entities by their English names. Output STRICT JSON
only - no markdown, no code fences, no commentary.

The recording is attached as audio input - listen to it first.

qa_format: {{qa_format}}
ambiguity_type: {{ambiguity_type}} - {{ambiguity_type_label}}
ambiguity_description: {{ambiguity_description}}

Spoken sentence, said twice (English): {{transcript_en}}
Reading 1 (English): {{interpretation_1_en}}
Reading 2 (English): {{interpretation_2_en}}

FIRST utterance (heard first)  GOLD reading: {{first_intended_en}}   (key phrase: {{first_gold_en}})
SECOND utterance (heard second) GOLD reading: {{second_intended_en}}  (key phrase: {{second_gold_en}})

Output JSON schema (use the one matching qa_format):
  mcq        -> {"question": str, "options": {"A": str, "B": str, ..., "J": str},   // exactly {{num_options}} options A-J
                 "answer": "A-J letter", "answer_text": str,
                 "GT_audio_reasoning": str,   // 5-6 sentence trace covering BOTH utterances in heard order
                 "rationale": str}
  open_ended -> {"question": str, "reference_answer_first": str, "reference_answer_second": str,
                 "GT_audio_reasoning": str,   // 5-6 sentence trace covering BOTH utterances in heard order
                 "rationale": str}

Return only the JSON object.
\end{tcblisting}

\begin{tcblisting}{listing only, breakable, enhanced,
  colback=black!3, colframe=black!55, boxrule=0.4pt, arc=1mm,
  left=2mm, right=2mm, top=1mm, bottom=1mm,
  title={Prosodic Ambiguity Resolution (V3): QA Generation Prompt}, fonttitle=\bfseries\small,
  listing options={basicstyle=\scriptsize\ttfamily, breaklines=true,
    columns=fullflexible, keepspaces=true}}
You are a meticulous benchmark author. A single audio clip contains the SAME Indonesian sentence
spoken TWICE by the same speaker. The words are identical both times; ONLY the prosody differs, so
each utterance expresses a DIFFERENT one of the sentence's two valid readings. THE CLIP ITSELF IS
ATTACHED as audio input - LISTEN to both utterances before writing anything. Your question asks
about EXACTLY ONE of the two utterances (the target), testing whether a listener can attend to that
specific utterance and resolve its meaning from the audio while ignoring the other.

The correct answer is FIXED by the TARGET utterance's GOLD reading - never infer it.

IMPORTANT - what the question may reveal: the question MUST begin with this exact sentence:
"The given audio has two different utterances."
Give NO other structural information: do NOT reveal that the two utterances are the same sentence,
and do not describe pauses, speakers, or recording details. After that fixed opening sentence, the
question MUST unambiguously identify the target as "the {{target_position_word}} utterance".

Follow the given qa_format:

- "mcq": After the fixed opening sentence, ask about the TARGET utterance, with exactly
  {{num_options}} concise options (letters A-J). This is a HARD benchmark: the option set must make
  every shortcut fail. Requirements:
    * Exactly ONE option matches the TARGET GOLD (the correct answer).
    * ONE option expresses the OTHER valid reading - the meaning carried by the NON-TARGET
      utterance. This is the hardest distractor: a listener who resolved the wrong utterance
      picks it.
    * ONE option is the safe-play escape: "The recording doesn't say" (or a natural equivalent).
      It must be WRONG - the target utterance does disambiguate.
    * Build the REMAINING options as the strongest wrong answers you can. HARD-DISTRACTOR rules:
        - prefer WRONG BINDINGS of entities/actions/locations that ARE in the sentence (swap
          which noun the property applies to, attach the phrase to the wrong action, invert
          who does what) - these force real parsing of the audio;
        - "both X and Y" / "neither X nor Y" mis-scopings;
        - subtle corruptions of the gold reading (right entity, wrong action; right action,
          wrong entity);
        - at most ONE option may introduce content not present in the sentence at all (besides
          the safe-play option) - foreign-content options are too easy to eliminate;
        - a distractor must NEVER amount to a third grammatically valid reading of the sentence
          - there must remain exactly one defensible answer.
    * DEDUPLICATION: every option must assert a DISTINCT state of affairs. No two options may
      differ only by a synonym, degree word (well/thoroughly), or rewording of the same claim.
      In particular, no two options may express the SAME underlying reading of the sentence in
      different words - each of the two valid readings appears exactly ONCE in the option set.
    * Every option must be a direct, well-formed answer to the question - no meta options.
    * Keep options concise, mutually exclusive, and similar in length/style so wording never
      leaks the answer.
    * Vary which letter is correct across items; do not default to a fixed position.
  Coordination / modifier scope (Type10) caution: one real reading already means "both ...", so never
  use "both" as a throwaway distractor there.

- "open_ended": After the fixed opening sentence, ask what the speaker means in the TARGET
  utterance. The question must be phrased so that EITHER reading's key phrase would be a natural
  and complete answer to it - it must not fit the target gold better than the other reading. Set
  reference_answer to the TARGET GOLD key phrase COPIED VERBATIM - do not paraphrase, expand, or
  reword it.

Question phrasing: apart from the mandatory opening sentence and the "{{target_position_word}}
utterance" reference, use natural, varied wording - no fixed template. Vary the phrasing style
across items.

CRITICAL - no transcript leakage: NEVER quote the sentence, reproduce it in translation, or closely
paraphrase its full content inside the question. Quoting it would also reveal that the two
utterances share the same words - structural information that must stay hidden. Refer only to the
minimal entities/events needed to pose the question. Additionally: do NOT quote ANY words or
phrases from the sentence (no quotation marks around its wording at all), and do NOT use
meta-linguistic pointers such as "the phrase X", "the prepositional phrase", or "the modifier" -
pose the question purely in terms of meaning. The question must never assert facts the listener
is supposed to extract by listening.

CRITICAL - frame neutrality (no answer bias): the question must be answerable under BOTH readings,
with DIFFERENT answers. It must NOT be built from the target gold's event frame, and the target
gold must not be guessable from the question text alone. Forbidden pattern: asking "where/when/how
does <the gold reading's action> happen?" when, under the other reading, that question would have
no answer or the answer "unspecified". SELF-CHECK before finalizing: silently answer your question
under reading 1 and under reading 2. If either reading yields no answer, "unspecified", or the
SAME answer as the other reading, REWRITE the question. This rule applies to BOTH mcq and
open_ended.

Coordination-scope phrasing (Type10): when the two readings differ on whether a property applies
to ONE item or BOTH items, the question must ask "WHICH item(s) ..." and let the listener choose
the subset. NEVER jointly predicate both items in the question (e.g. "what is said about the
condition of A and B?" is FORBIDDEN - it fits the both-reading better; ask "which of the stolen
items are described as new?" style instead).

Format example (ILLUSTRATION ONLY - do not reuse its wording, topic, sentence pattern, or option
style; your item must be built from the sentence you are given):
  Question: "The given audio has two different utterances. In the first utterance, what did the
  speaker personally pick up at the beach?"
  Options:  A) The sand   B) The shell   C) Both the sand and the shell
            D) The recording doesn't say   ... (continues through J with strong wrong answers)

REQUIRED FIELD - GT_audio_reasoning (both formats): after listening to the attached clip, write a
5-6 sentence ground-truth audio reasoning trace grounded in THIS specific clip: (1) note the clip
contains two utterances of the same words and that the question targets the
{{target_position_word}} one; (2) describe the concrete acoustic cues you actually perceive in the
TARGET utterance (prosodic boundary/pause placement, stressed or lengthened words, phrase
grouping, pacing) and how they select the TARGET GOLD reading; (3) briefly contrast with the other
utterance's delivery so the difference is explicit. Describe ONLY cues you can genuinely perceive
- NEVER invent or assume cues; if the audio does not clearly support the target gold reading,
state that explicitly inside the trace. Do not reference option letters - the trace must stand
alone as the reference reasoning for this clip.

Write in {{question_language}}, referring to entities by their English names. Output STRICT JSON
only - no markdown, no code fences, no commentary.

The recording is attached as audio input - listen to it first.

qa_format: {{qa_format}}
ambiguity_type: {{ambiguity_type}} - {{ambiguity_type_label}}
ambiguity_description: {{ambiguity_description}}

Spoken sentence, said twice (English): {{transcript_en}}
Reading 1 (English): {{interpretation_1_en}}
Reading 2 (English): {{interpretation_2_en}}

TARGET utterance: the {{target_position_word}} utterance in the clip (position {{target_position}}).
TARGET GOLD reading (CORRECT answer): {{target_intended_en}}   (key phrase: {{target_gold_en}})
The OTHER (non-target) reading: {{other_reading_en}}

Output JSON schema (use the one matching qa_format):
  mcq        -> {"question": str, "options": {"A": str, "B": str, ..., "J": str},   // exactly {{num_options}} options A-J
                 "answer": "A-J letter", "answer_text": str, "distractor_letters": [str, ...],
                 "GT_audio_reasoning": str,   // 5-6 sentence audio reasoning trace from THIS recording
                 "rationale": str}
  open_ended -> {"question": str, "reference_answer": str,
                 "GT_audio_reasoning": str,   // 5-6 sentence audio reasoning trace from THIS recording
                 "rationale": str}

{{type10_reminder}}Return only the JSON object.
\end{tcblisting}

\begin{tcblisting}{listing only, breakable, enhanced,
  colback=black!3, colframe=black!55, boxrule=0.4pt, arc=1mm,
  left=2mm, right=2mm, top=1mm, bottom=1mm,
  title={Dialectal Speech Comprehension: QA Generation Prompt}, fonttitle=\bfseries\small,
  listing options={basicstyle=\scriptsize\ttfamily, breaklines=true,
    columns=fullflexible, keepspaces=true}}
You are a senior speech-benchmark editor with native-level command of Thai (including
Northern/Khummuang, Northeastern/Korat, and Southern Pattani - Pattani is Pattani Malay) and
Vietnamese (all regional accents). You receive ONE benchmark item: the audio clip, its
transcript(s), and its QA (question + gold answer, plus 10 options for MCQ or answer aliases
for open-ended). The evaluated models will only ever hear the AUDIO - never the transcript.

CORE GOAL: enhance the quality of this existing item so that it is EXTREMELY HARD for an
audio model that must genuinely listen - while staying perfectly fair (exactly one defensible
answer, fully supported by the audio). A hard item forces real listening; it never wins by
being ambiguous, unanswerable, or by tricking the grader.

================ MANDATORY ANALYSIS PROTOCOL - follow IN ORDER before writing ================

STEP 1 - LISTEN to the attached audio end-to-end. Actually hear it: the dialect/accent, every
content word, every name and number, hesitations, and anything hard to perceive. Note where a
non-native or standard-variety listener would mishear.

STEP 2 - READ the dialect transcript as reference material: it tells you what is spoken so
you can interpret the dialect content precisely. The standard-Thai line and English gloss
are unreliable corpus translations - never treat them as ground truth. Do NOT audit,
report, or comment on audio<->transcript alignment (word-by-word presence, extra/missing
words, etc.) - that is NOT the task. The task is only about the QA: is the question
answerable from the audio, is the gold correct and unique, and can the item be made harder.

STEP 3 - READ the question. Understand exactly what it asks and which part of the audio
answers it.

STEP 4 - READ the answer side. MCQ: read all 10 options A-J as a set - which are strong,
which are weak, which could a listener argue for? Open-ended: read the gold answer and every
alias - would any alias accept a wrong answer?

STEP 5 - FEASIBILITY CHECK. Decide and report `feasible`:
  * Is the question answerable from the audio ALONE (no outside knowledge, no transcript)?
  * Is the gold unambiguously correct per the audio?
  * Is it the ONLY defensible answer - no other option (A-H) and no alternative reading of
    the audio is arguably correct?
  * MCQ: options mutually exclusive, same type, same style? Open-ended: aliases tight?
  If any of these fail and you cannot repair them within the contract below, set
  `feasible` = false and explain in `issues_found`.

STEP 6 - LEAKAGE AUDIT + HARDENING (this is the core editorial pass; be aggressive).
The question must be VERY GENERAL: a reader who only sees the question (and options) must
learn NOTHING about the answer. Audit the existing question for leakage and FIX every leak:
  * it names, quotes, or distinctively paraphrases the answer, its value, unit, or entity;
  * it quotes dialect words or any transcript wording, or summarizes the clip's content as
    scaffolding/context before asking;
  * it narrows the field unfairly (mentions a category, count, place, or timeframe that only
    the gold fits, or that eliminates distractors without listening);
  * it asserts facts the listener is supposed to extract by listening;
  * MCQ: the gold option is identifiable without audio - by length, grammar fit with the
    question, style mismatch with the distractors, or by being the only same-type answer.
Then HARDEN wherever the audio supports it (never make the item easier):
  * question: generic wording in the style "According to the audio, ..." - answerable from
    audio alone, in English, no template monotony, zero leakage;
  * MCQ distractors: replace weak ones with plausible mishearings of THIS dialect/accented
    audio and with competing facts genuinely audible in the clip (other numbers, names, or
    items the speaker really says) - same type, similar length, lowercase-initial;
  * open-ended: keep 4-8 tight aliases (digits AND words for numbers; native-script dialect
    + standard forms where relevant); never an alias so broad it matches a wrong answer.
Surgical edits only - improve the existing item; do not invent a different question about a
different part of the clip unless the current target is unrecoverable (then say so in
`issues_found` and set `feasible` accordingly).

BENCHMARK CONTRACT - never break these:
  * MCQ has EXACTLY 10 options labelled A-J. Two safe-guess trap options - "{{trap_i}}" and
    "{{trap_j}}" - appear somewhere among A-J; their letters VARY per item (given as
    `trap_letters` in the payload). Both are ALWAYS incorrect by design - never make them
    the answer, never move them to other letters, never reword them.
  * The gold answer MUST stay at letter {{gold_letter_rule}}.
  * Every other option keeps its letter - you may improve a distractor's TEXT, never its
    position.
  * If the item is already correct, leak-free, and maximally hard, set verdict "unchanged"
    and copy the fields as-is.

STEP 7 - GROUND-TRUTH AUDIO-REASONING TRACE (always produce this, for every item).
Write `GT_audio_reasoning`: ONE flowing prose paragraph of 3-5 sentences - the genuine
reasoning of an expert listener working out the answer to this question from the audio. You
produce it AFTER analyzing everything, but you WRITE it as pure listening reasoning. It is
about HOW TO ANSWER the question, never about how the item was constructed. It is the
reference trace that evaluated models' reasoning will be compared against. Requirements:
  * NO numbered or bulleted steps - a single analytic paragraph.
  * NO scene-setting boilerplate ("The audio is spoken in Pattani Malay...", "The clip is
    about..."): start directly from the evidence that decides the question - the key
    word/phrase AS HEARD (quote the dialect/accented form, native script where useful),
    what it means, why a naive listener could mis-map or mishear it, why the strongest
    competing candidate fails, and the conclusion stated as the final answer.
  * Describe ONLY what is genuinely audible in THIS clip - never invent or assume acoustic
    cues. If the audio does not clearly support the gold, note that in `issues_found`.
  * Never mention transcripts, glosses, datasets, option letters, or that you were given any
    text. Refer to wrong candidates by their content, not their letter. The paragraph must
    stand alone ("The speaker says ... which in Northern Thai means ...").

================================== OUTPUT ==================================
Return ONE strict JSON object, no markdown fences, no commentary:
{
  "id": "<same id>",
  "feasible": true|false,
  "leakage_found": ["<each leak found in the ORIGINAL question/options; empty if none>"],
  "issues_found": ["<short strings, empty list if none>"],
  "verdict": "unchanged" | "revised",
  "question": "<final question>",
  "gold_answer": "<final gold>",
  "options": {"A": "...", ..., "J": "..."},        // MCQ only, all 10, traps intact at their letters
  "answer_aliases": ["..."],                        // open-ended only
  "GT_audio_reasoning": "<one prose paragraph, 3-5 sentences>",
  "change_notes": "<=25 words; empty string if unchanged"
}

The audio clip is attached in this message - LISTEN TO IT FIRST, then follow the protocol
(audio -> transcript -> question -> options/aliases -> feasibility -> harden/de-leak -> trace).

Benchmark item to verify, improve, and trace:
{{payload_json}}

Return ONLY the JSON object.
\end{tcblisting}

\begin{tcblisting}{listing only, breakable, enhanced,
  colback=black!3, colframe=black!55, boxrule=0.4pt, arc=1mm,
  left=2mm, right=2mm, top=1mm, bottom=1mm,
  title={Dialect and Language Identification: Open-Ended Judge Prompt}, fonttitle=\bfseries\small,
  listing options={basicstyle=\scriptsize\ttfamily, breaklines=true,
    columns=fullflexible, keepspaces=true}}
You are an expert QA evaluator for a spoken DIALECT / Language identification
task. You are given the QUESTION, the REFERENCE (correct) answer(s), and a
model's free-text answer about the same audio clip. Evaluate the model's
response against the reference answer, which is the GROUND TRUTH.

QUESTION: {question}
REFERENCE ANSWER (Ground Truth): {reference_answer}
MODEL RESPONSE: {model_response}

EVALUATION CRITERIA (Score 0 or 1):
- Score 1 if the model response conveys the SAME meaning as ANY reference
  answer, even with different wording or order.
- The answer may be BRIEF: it need NOT restate information already given in
  the question (e.g. Q "which segment is Javanese?" -> "segment 3" scores 1).
- Ignore case, punctuation, filler words, politeness, and trivial
  singular/plural differences that do not change the meaning.
- If the reference lists MULTIPLE items or an ORDERED sequence, score 1 only
  if the model conveys ALL of them in the correct order; partial or
  out-of-order coverage scores 0.
- Score 0 if the model gives a different dialect, segment, order, count, or
  scope than the reference, or is vague, hedging, or empty.
- Return ONLY valid JSON (no markdown, no extra text).

Output should be in the following format:

{{
  "score": 0 or 1,
  "justification": "1-2 sentences about why you gave that score"
}}
\end{tcblisting}

\begin{tcblisting}{listing only, breakable, enhanced,
  colback=black!3, colframe=black!55, boxrule=0.4pt, arc=1mm,
  left=2mm, right=2mm, top=1mm, bottom=1mm,
  title={Prosodic Ambiguity Resolution: Open-Ended Judge Prompt}, fonttitle=\bfseries\small,
  listing options={basicstyle=\scriptsize\ttfamily, breaklines=true,
    columns=fullflexible, keepspaces=true}}
You are an expert QA evaluator for a spoken-language DISAMBIGUATION task.
The underlying sentence is structurally ambiguous and has two possible
readings; the REFERENCE answer states the single intended reading. You are
given the QUESTION, the REFERENCE (correct) answer(s), and a model's
free-text answer. Evaluate the model's response against the reference, which
is the GROUND TRUTH.

QUESTION: {question}
REFERENCE ANSWER (Ground Truth): {reference_answer}
MODEL RESPONSE: {model_response}

EVALUATION CRITERIA (Score 0 or 1):

A. WHAT COUNTS AS CORRECT (score 1): the model conveys the SAME reading as
   the reference. The answer MAY be brief and need NOT restate information
   already given in the question. If the question asks WHICH entity, WHERE,
   WHEN, or HOW MANY, naming the correct entity or value is a COMPLETE
   answer by itself.
   - Example: Q "which item is blue?", reference "the folder is blue",
     response "the folder" -> score 1.
   - Example: Q "where is the bell?", reference "the bell is in front of the
     class", response "in front of the class" (or "rang the bell in front of
     the class") -> score 1; the required location is present.
   Paraphrases and different wording with the same meaning score 1.

Score 0 if ANY of the following apply:
B. NOT COMMITTED: the answer is vague, generic, or compatible with BOTH
   readings - e.g. restating the ambiguous sentence, "they are dead", "it
   was expensive", "either", "cannot tell", "the audio doesn't specify", or
   presenting both readings without choosing.
C. WRONG SCOPE: the reference says BOTH items have the property but the model
   asserts only one; OR the reference restricts it to ONE item but the model
   says "both" or leaves it unrestricted.
D. WRONG ORDER (two-utterance answers): the model reverses the
   first-utterance and second-utterance readings relative to the reference.
E. WRONG CONTENT: the model states a different entity, action, location, or
   scope than the reference.

Ignore case, punctuation, filler words, politeness, and trivial
singular/plural differences that do not change the meaning.

Return ONLY valid JSON (no markdown, no extra text). Output format:

{{
  "score": 0 or 1,
  "justification": "1-2 sentences about why you gave that score"
}}
\end{tcblisting}

\begin{tcblisting}{listing only, breakable, enhanced,
  colback=black!3, colframe=black!55, boxrule=0.4pt, arc=1mm,
  left=2mm, right=2mm, top=1mm, bottom=1mm,
  title={Dialectal Speech Comprehension: Open-Ended Judge Prompt}, fonttitle=\bfseries\small,
  listing options={basicstyle=\scriptsize\ttfamily, breaklines=true,
    columns=fullflexible, keepspaces=true}}
You are an expert, MULTILINGUAL QA evaluator for a DIALECTAL SPEECH
COMPREHENSION task. A speaker talks in a regional dialect; the model must
report what was said. You are given the QUESTION, the REFERENCE (correct)
answer(s), and a model's free-text answer about the same audio clip.
Evaluate the model's response against the reference, which is the GROUND
TRUTH. The reference may be written in English; the model's answer may be in
English, in the native language/script (e.g. Thai or Vietnamese), or in
transliteration.

QUESTION: {question}
REFERENCE ANSWER (Ground Truth): {reference_answer}
MODEL RESPONSE: {model_response}

EVALUATION CRITERIA (Score 0 or 1):
- LANGUAGE-AGNOSTIC: Judge the MEANING / REFERENT, never the language or
  script. Score 1 if the model answer denotes the SAME entity, place,
  person, time, amount, or fact as the reference - whether written in
  English, in the native script, or transliterated. A correct native-script
  or translated answer scores 1.
- NAME / TRANSLITERATION VARIANTS: Accept minor spelling, diacritic, tone-
  mark, spacing, or transliteration differences (including one- or two-
  character differences) in a name or place when they clearly refer to the
  SAME referent (e.g. "Nong Khu" = "Nong Khru", including the same name
  written in Thai script; "Li Peng Ma Yeng" = "Lipeng Mayeng"). Do NOT
  accept a genuinely different name, place, number, or entity.
- EQUIVALENT FORMS: Accept the same date/number written differently
  (e.g. "June 30, 2021" = "30/6/2021"), and paraphrases with the same
  meaning. The answer may be BRIEF and need not restate the question.
  Ignore case, punctuation, filler, and politeness.
- MULTIPLE PARTS: If the reference requires MULTIPLE parts, score 1 only if
  the model conveys ALL of them.
- Score 0 if the model gives a genuinely different entity/place/time/amount/
  fact, or is vague, hedging ("cannot tell", "not specified"), or empty. Do
  NOT give credit for merely restating the question or the audio.
- Return ONLY valid JSON (no markdown, no extra text).

Output should be in the following format:

{{
  "score": 0 or 1,
  "justification": "1-2 sentences about why you gave that score"
}}
\end{tcblisting}

\begin{tcblisting}{listing only, breakable, enhanced,
  colback=black!3, colframe=black!55, boxrule=0.4pt, arc=1mm,
  left=2mm, right=2mm, top=1mm, bottom=1mm,
  title={Long-Form Audio Reasoning: Open-Ended Judge Prompt}, fonttitle=\bfseries\small,
  listing options={basicstyle=\scriptsize\ttfamily, breaklines=true,
    columns=fullflexible, keepspaces=true}}
You are an expert QA evaluator. Evaluate the model's response against the reference answer, which is the GROUND TRUTH.

QUESTION: {question}
REFERENCE ANSWER (Ground Truth): {reference_answer}
MODEL RESPONSE: {model_response}

STRICT EVALUATION CRITERIA (Score 0.0-1.0). Grade on SUBSTANCE - whether the answer is
factually right, on-topic, and complete. Do NOT reward fluent writing or confident tone
on its own; a well-written but incorrect answer must still score low.

1. CORRECTNESS: Does the model's response match the factual content of the reference answer?
   - 1.0: Factually equivalent to reference (unambiguous paraphrases/synonyms count)
   - 0.5-0.9: Mostly correct with minor discrepancies
   - 0.0-0.4: Contains a factual error or contradicts the reference

2. RELEVANCE: Does the response directly address the question asked, without evasion?
   - 1.0: Directly answers the question
   - 0.5-0.9: Mostly on-topic with minor tangents
   - 0.0-0.4: Restates the question, hedges, or answers a different question

3. COMPLETENESS: Does the response cover ALL key points the reference answer requires?
   - 1.0: Every required point present
   - 0.5-0.9: A minor detail missing
   - 0.0-0.4: A major required point missing

STRICT SCORING RULES:
- average_score = 0.6*correctness + 0.1*relevance + 0.3*completeness
  (correctness and completeness dominate; do NOT flat-average the criteria).
- OMISSION CAP: if the answer omits any required part of the reference answer,
  average_score <= 0.60 (a partially complete answer cannot score high).
- No partial credit for restating the question, hedging, or listing possibilities
  without committing to a single answer.
- Full credit (>=0.9) requires ALL required facts, stated correctly. "Mostly right"
  is not full credit.

Return ONLY valid JSON (no markdown, no extra text):
{{
  "correctness": {{"score": 0.0-1.0, "justification": "1-2 sentences"}},
  "relevance": {{"score": 0.0-1.0, "justification": "1-2 sentences"}},
  "completeness": {{"score": 0.0-1.0, "justification": "1-2 sentences"}},
  "overall": {{
    "average_score": 0.0-1.0,
    "overall_assessment": "2-3 sentence summary; note if the omission cap was applied"
  }}
}}
\end{tcblisting}

\begin{tcblisting}{listing only, breakable, enhanced,
  colback=black!3, colframe=black!55, boxrule=0.4pt, arc=1mm,
  left=2mm, right=2mm, top=1mm, bottom=1mm,
  title={Speech Emotion Recognition: Open-Ended Judge Prompt}, fonttitle=\bfseries\small,
  listing options={basicstyle=\scriptsize\ttfamily, breaklines=true,
    columns=fullflexible, keepspaces=true}}
You are an expert QA evaluator for speech emotion-recognition task. Evaluate the model's response against the reference answer which is the GROUND TRUTH for the question.

QUESTION: {question}
REFERENCE ANSWER (Ground Truth): {reference_answer}
MODEL RESPONSE: {model_response}

RELAXED EVALUATION CRITERIA (Score 0 or 1):
- Score 1 if the model's free-text answer correctly identifies the correct emotion.
- Score 0 otherwise (wrong emotion, no commitment, or only a vague/contradictory description).
- Accept the model response if it conveys the same core concept, or conclusion, even if phrased differently or structured differently.
- Accept minor omissions of secondary detail as long as the primary reasoning is correct and not contradicted.
- Award a score of 0 if the model response contains a significant factual error, contradicts the reference answer's core reasoning, or introduces misleading information.
- Award a score of 0 if the model response is vague or generic to the point of not addressing the specific context implied by the question.
- Return ONLY valid JSON (no markdown, no extra text).

Output should be in the following format:

{{
  "score": 0 or 1,
  "justification": "1-2 sentences about your given score like why you have given that score"
}}
\end{tcblisting}

\begin{tcblisting}{listing only, breakable, enhanced,
  colback=black!3, colframe=black!55, boxrule=0.4pt, arc=1mm,
  left=2mm, right=2mm, top=1mm, bottom=1mm,
  title={Speech-Affective Interpretation: Open-Ended Judge Prompt}, fonttitle=\bfseries\small,
  listing options={basicstyle=\scriptsize\ttfamily, breaklines=true,
    columns=fullflexible, keepspaces=true}}
You are an expert QA evaluator for speech affective-Interpretation task with two axes:
VALENCE / mood (Positive, Negative, Neutral) and AROUSAL / energy (High, Low, Neutral). 

You are given the CORRECT "valence, arousal" answer and a model's free-text answer about the same audio clip. Evaluate the model's response against the reference answer which is the GROUND TRUTH for the question.

QUESTION: {question}
REFERENCE ANSWER (Ground Truth): {reference_answer}
MODEL RESPONSE: {model_response}

RELAXED EVALUATION CRITERIA (Score 0 to 1):
- Score 1 ONLY if the model's free-text answer correctly conveys BOTH the correct valence AND the correct arousal (unambiguous synonyms count: e.g. pleasant=Positive, upset/down=Negative, calm/flat=Neutral; energetic/agitated=High, subdued/quiet=Low, moderate/even=Neutral).
- Score 0 if either axis is wrong, missing, or only vaguely implied.
- Award a score of 0 if the model response contains a significant factual error, contradicts the reference answer's core reasoning, or introduces misleading information.
- Award a score of 0 if the model response is vague or generic to the point of not addressing the specific context implied by the question.
- Return ONLY valid JSON (no markdown, no extra text).

Output should be in the following format:

{{
  "score": 0 or 1,
  "justification": "1-2 sentences about your given score like why you have given that score"
}}
\end{tcblisting}

\begin{tcblisting}{listing only, breakable, enhanced,
  colback=black!3, colframe=black!55, boxrule=0.4pt, arc=1mm,
  left=2mm, right=2mm, top=1mm, bottom=1mm,
  title={Reasoning-Trace Audit: Classification Prompt}, fonttitle=\bfseries\small,
  listing options={basicstyle=\scriptsize\ttfamily, breaklines=true,
    columns=fullflexible, keepspaces=true}}
You are auditing the stated reasoning of an audio language model. For one
benchmark record you receive: the QUESTION (and options, if multiple-choice),
the GOLD answer, the REFERENCE transcript and gold metadata for the audio,
the model's PREDICTED answer, and the model's STATED REASONING. Judge only
the stated reasoning against the reference material; do NOT re-answer the
question yourself.

Label the record on two axes.

AXIS 1 - GROUNDING (label every record):
- grounded      : the reasoning cites specific audible content, consistent
                  with the reference transcript, that supports the model's
                  own answer.
- partial       : some genuine audio evidence, padded with unsupported leaps.
- ungrounded    : generic or templated justification, pure option
                  elimination, or answer restatement with no audio evidence.
- contradictory : the stated reasoning points to a DIFFERENT answer than the
                  one the model gave.

AXIS 2 - FAILURE TYPE (label only records whose predicted answer is wrong):
- perception    : mis-heard words or numbers.
- comprehension : heard approximately right but mapped to the wrong meaning.
- hallucination : invents content absent from the audio.
- reasoning     : evidence right, derivation wrong.
- abstention    : refused or hedged on an answerable item.
- format        : invalid or empty output.

Return ONLY valid JSON (no markdown, no extra text):

{{
  "grounding": "grounded | partial | ungrounded | contradictory",
  "failure_type": "perception | comprehension | hallucination | reasoning | abstention | format | null",
  "rationale": "1-2 sentences citing the decisive evidence"
}}
\end{tcblisting}

\end{document}